\documentclass[a4paper, 11pt]{article}
\pdfoutput=1

\usepackage{jcappub}

\usepackage{graphicx}
\usepackage{amsmath}
\usepackage{amsfonts}
\usepackage{amssymb}
\usepackage{bbold}
\usepackage{float}

\usepackage{tikz}
\usetikzlibrary{arrows,backgrounds,snakes,patterns}
\usetikzlibrary{shapes,arrows,chains}
\usepackage{verbatim}
\usepackage{booktabs}
\usepackage{pgfplots}
\pgfplotsset{compat=1.10}
\usepgfplotslibrary{fillbetween}
\usepackage[flushleft]{threeparttable}

\usepackage{array}
\usepackage{placeins}

\newcommand{\Tr}{\operatorname{Tr}}

\newcommand{\Mpl}{\ensuremath M_{\textrm{pl}{}}}

\numberwithin{equation}{section}
\numberwithin{figure}{section}

\dedicated{UTTG-10-18}
\title{ Inflation in Multi-field Modified DBM Potentials}

\author[a]{Sonia Paban}
\author[a]{Robert Rosati}

\affiliation[a]{
	Theory Group, Department of Physics, University of Texas at Austin, Austin, TX 78712, USA
}

\emailAdd{paban@physics.utexas.edu}
\emailAdd{rjrosati@utexas.edu}

\abstract{We study multi-field inflation in random potentials generated via a non-equilibrium random matrix theory process. We make a novel modification of the process to include correlations between the elements of the Hessian and the height of the potential, similar to a Random Gaussian Field (RGF).
We present the results of over 50,000 inflationary simulations involving 5-100 fields. For the first time, we present results of $\mathcal{O}(100)$ fields using the full `transport method', without slow-roll approximation. We conclude that Planck compatibility is a common prediction of such models, however significant isocurvature power at the end of inflation is possible.}

\arxivnumber{1807.07654}

\begin{document}
	
\maketitle
\thispagestyle{empty}
\newpage

\setcounter{page}{1}

\section{Introduction}
\par Planck \cite{Ade:2015lrj} put tight constraints on the observation of isocurvature modes. Theoretically, it is well known that multi-field inflation models, ubiquitous in UV complete theories,  can generate isocurvature modes  but can also evade Planck's limits if these modes are washed out by later processes \cite{Mollerach:1989hu,Weinberg:2004kf,Hotinli:2017vhx}. Since our understanding of the reheating period is not firm, there has been a substantial effort toward clarifying the conditions that lead to the creation of isocurvature modes \cite{Lyth:1984gv,Starobinsky:1986fxa,Salopek:1990jq,Sasaki:1995aw,Sasaki:1998ug,GrootNibbelink:2000vx,GrootNibbelink:2001qt,Wands:2000dp,Amendola:2001ni,Rigopoulos:2004gr,Rigopoulos:2005xx,Yokoyama:2007uu,Lalak:2007vi, Ringeval:2007am, Yokoyama:2008by,Malik:2008im,Mulryne:2009kh,Chen:2009zp, Mulryne:2010rp,Peterson:2010np,Peterson:2011yt,Achucarro:2010da,Cremonini:2010ua,Avgoustidis:2011em, Lehners:2009ja,Huston:2009ac,Kaiser:2013sna,Schutz:2013fua, Dias:2016slx,Dias:2017gva, Masoumi:2016eag, Masoumi:2017gmh, Masoumi:2017xbe, Freivogel:2016kxc,Bjorkmo:2017nzd, Bjorkmo:2018txh, Achucarro:2017ing}. This work adds to this body of research by further exploring the generation of isocurvature modes in a class of random potentials generated through the Dyson Brownian Method (DBM).

In any multi-dimensional space a trajectory is always one-dimensional, so it is always possible to change variables so that only one of them is excited along the path.
This is not true for the quantum fluctuations \cite{GrootNibbelink:2000vx, GrootNibbelink:2001qt, Malik:2008im, Achucarro:2010da, Cremonini:2010ua, Schutz:2013fua}, that can be decomposed in fluctuations parallel to the classical path (adiabatic mode) and perpendicular (isocurvature modes).
After  horizon exit, isocurvature modes source the adiabatic mode proportionally to the turning rate of the classical trajectory.
The isocurvature perturbations, in turn, source each other, have an amplitude that decays exponentially if $0 \leq m^2 \gg H^2$ (see \cite{Chen:2009zp} for $m^2 \sim H^2$) but have the potential to grow temporarily when slow-roll conditions don't hold \cite{Leach:2001zf}.


Extracting useful information about the effect of light fields on the evolution is made harder by the lack of a preferred potential for UV-complete inflation. One approach is to tackle the problem from the bottom up, and try to find universal traits in families of potentials.
One popular family of potentials is the Random Gaussian Field (RGF) \footnote{Often, the central limit theorem is given as the justification for this choice. Though a potential in the multi-field ($N \sim 100)$ limit is the sum of a large number of terms, one of the requirements of the theorem, there is no reason to expect the different terms to be equally distributed, an assumption that would guarantee the theorem is satisfied.}. This family of potentials is characterized by the following relations:
\begin{eqnarray}
\langle V({\pmb \phi}) \rangle &=&  \bar{V} \\
\langle  V({\pmb \phi}_1 )     V({\pmb \phi}_2 )   \rangle-\bar{V}^2 &=& V_0^2 \exp{ \left(    - | {\pmb \phi}_1-{\pmb \phi}_2|^2 /{2 \Lambda_h^2}            \right)}   
\label{eq:RGFcorr}
\end{eqnarray} where $\bar{V}$, $V_0$ and $\Lambda_h$ are constants.

Simulating RGFs is numerically costly. Chronologically one of the first attempts was due to Marsh, McAllister, Pajer and Wrase \cite{Marsh:2013qca} who proposed a  method that charts the evolution of the potential near and along the classical trajectory by stipulating that the Hessian matrices in adjacent coordinate patches are related by Dyson Brownian Motion (DBM) (a detailed explanation of the method can be found  in Section \ref{sec:potentials}). Later, a more comprehensive numerical analysis by Dias, Frazer and Marsh \cite{Dias:2016slx,Dias:2017gva}, using this method, found that Planck compatibility is common, both in the results for $n_s$ and in the size of the isocurvature power (in fact, although significant isocurvature power is generated, it can decay). It is unclear, however, to what extent, if any, the potentials generated through this method are RGFs. At a given point, a true RGF, described by (\ref{eq:RGFcorr}), has correlations between all even-ordered Taylor coefficients, and all odd-ordered Taylor coefficients while 
the odd and even coefficients are uncorrelated \cite{Yamada:2017uzq,Masoumi:2016eag,Masoumi:2017gmh}. The DBM approach only constraints the Hessian coefficients. There are other reasons to suspect this connection as detailed in \cite{Marsh:2013qca, Bachlechner:2014rqa, Freivogel:2016kxc, Masoumi:2016eag,Masoumi:2017gmh,Bjorkmo:2017nzd} and discussed in Section \ref{sec:potentials}.  Masoumi, Vilenkin and Yamada \cite{Masoumi:2016eag,Masoumi:2017gmh} use only (\ref{eq:RGFcorr}) to derive statistical results on inflection-point inflation. The potential is generated locally around an inflection- point, and (\ref{eq:RGFcorr}) is used to constraint the Taylor coefficients up to cubic order. 
For inflation to proceed  the masses around such a point should be positive or zero which, given the average mass separation derived from (\ref{eq:RGFcorr}) and the small field excursion, justifies a single-field behavior.  This argument is likely to be true in most cases, but not always (as the masses are drawn from a distribution there will be instances when they are  sufficiently close to invalidate the argument.)  Bjorkmo and Marsh \cite{Bjorkmo:2017nzd, Bjorkmo:2018txh}, improve on an early suggestion by \cite{Bachlechner:2014rqa}, and generate an approximate saddle point RGF inflationary potential locally around the saddle point, with $\Lambda_h < M_P$ and $N_f \geq 6$. Their work examines many realizations of these potentials and finds that despite substantial multi-field effects, Planck compatibility is not rare.

These approaches can be divided into two groups.  In \cite{Masoumi:2016eag,Masoumi:2017gmh, Bjorkmo:2017nzd, Bjorkmo:2018txh} the potential around the inflection or saddle point is smooth and given by a polynomial potential with random coefficients following (\ref{eq:RGFcorr}). Though a polynomial can't represent a RGF (it is not translationally invariant), the short-field excursion during inflation justifies its use. In the DBM method the motion of the Hessian from point to point is supposed to account for the higher order terms in the Taylor expansion. The Hessian evolution is continuous though non-differentiable and is determined by the DBM dynamics as explained in section \ref{sec:potentials}. This is an additional input.  In this work we revisit the issue of multi-field inflation with an improved DBM that takes into account the correlation between the Hessian and the value of the potential. With this correlation the DBM dynamics gives the flow from the non-generic point where inflation starts to a generic critical point. This new method generates another family of potentials against which to check the robustness of the predictions made by the approaches of \cite{Masoumi:2016eag,Masoumi:2017gmh, Bjorkmo:2017nzd, Bjorkmo:2018txh} as we vary the parameters $N_f, \Lambda_h$ and $\bar{V}$. 
We use an initial mass distribution that is consistent with (\ref{eq:RGFcorr}) and inflation \cite{Yamada:2017uzq}, and is different from that used in  \cite{Dias:2016slx,Dias:2017gva}, because \cite{Yamada:2017uzq} is derived from a RGF.
To compute the evolution of the power spectrum we use a version of the transport method \cite{Dias:2015rca} that does not assume slow-roll given that the slow-roll conditions aren't always satisfied along inflation. 

\section{Potential Construction}
\label{sec:potentials}
In the DBM method we write the potential near a point in field space $p_i$ as a quadratic Taylor expansion:
\begin{align}
V(\vec{\phi}) = \Lambda_{\textrm v}^4 \sqrt{N_f} (v_0 + v_I \tilde\phi^I + v_{IJ}\tilde{\phi}^I \tilde{\phi}^J)
\label{eq:potential}
\end{align}
where the $\tilde{\phi}^I \equiv \phi^I / \Lambda_h $ are the $\mathcal{O}(100)$ fields normalized by the RGF correlation length, $\Lambda_h$.
The matrix, $v_{IJ}$, performs a random walk via Dyson Brownian Motion (DBM) as the fields evolve.
Each timestep of the evolution, the potential becomes a new quadratic approximation centered at the new position of the fields in field space.
Precisely,
\begin{align}
    \begin{split}
    v_{IJ}|_{p_{i+1}} &= v_{IJ}|_{p_i} + \delta v_{IJ}|_{p_i\rightarrow p_{i+1}} \\
    v_I |_{p_{i+1}}   &= v_I|_{p_i} + v_{IJ}|_{p_{i}} \delta \tilde{\phi}^J \\
    v_0 |_{p_{i+1}}   &= v_0 |_{p_{i}} + v_I |_{p_{i}} \delta \tilde{\phi}^I
    \end{split}
    \label{eq:DBMcoeffs}
\end{align}
where the $\delta \tilde{\phi}^I \equiv \delta \phi^I / \Lambda_h$ is the field displacement given by the fields' equation of motion in cosmic time interval $\delta t$ and the $\delta v_{IJ}$ are independent and identically distributed random variables with fixed ensemble averages for $\langle \delta v_{IJ}\rangle$ and $\langle \delta v_{IJ}^2\rangle$.
In DBM all other averages are of higher order in $\delta s \equiv \sqrt{\delta \tilde{\phi}_I \delta \tilde{\phi}^I}$.
In \cite{Marsh:2013qca,Dias:2016slx,Dias:2017gva} these were taken to be:

\begin{align}
    \begin{split}
	\langle \delta v_{IJ} \rangle &= \left( -v_{IJ} \right) \delta s \\ \langle \delta v_{IJ}^2 \rangle &= \sigma^2 \left(1+\delta_{IJ} \right) \delta s
    \end{split} 
    \label{eq:DBMpdf}
\end{align}
where $\sigma^2$ is a constant.
This evolution, as explained in \cite{Mehta1991,Mehta2004,doi:10.1063/1.1703862}, is equivalent to solving the Fokker-Planck-Smoluchowski equation for the probability density function (PDF) of $v_{IJ}$, ${\cal  P}(v_{IJ},s)$,
\begin{align}
    \frac{\partial {\cal P}}{\partial s}= \sum_{I=1}^{N_f}\sum_{J=1}^{I} \left(\frac{1}{2} \sigma^2 (1+\delta_{IJ}) \frac{\partial^2 \cal P}{\partial v_{IJ}^2} + \frac{\partial} {\partial v_{IJ}}(v_{IJ} {\cal P} ) \right) 
\end{align}
The stationary solution of this equation is the PDF for a matrix in the Gaussian Orthogonal Ensemble (GOE):
\begin{align}
   {\cal  P}(v) \propto e^{-  \Tr{v^2}/(2 \sigma^2)}.
\end{align} 

As mentioned earlier the DBM proposal to generate random potentials \cite{Marsh:2013qca,Freivogel:2016kxc,Dias:2016slx} is not without complications. Indeed several authors\cite{Bachlechner:2014rqa,Easther:2016ire,Masoumi:2016eag,Wang:2016kzp,Masoumi:2017xbe} have pointed out a series of shortcomings derived from the observation that the potentials generated through this method: (i)  are most likely unbounded, (ii) are not smooth functions, (iii) are not single-valued, (iv) do not include correlations between the elements of the Hessian and (v) have too few maxima and minima relative to saddle points, compared to Morse functions.

These objections are all valid, but they weigh differently for the study of multi-field inflation. In this work we modify (\ref{eq:DBMpdf}) to incorporate the correlation between the height of the potential and the Hessian. As the potential gets higher (lower) than its mean, each critical point becomes more likely to be a local maximum (minimum). This, unfortunately, does not bound our potentials, but as explained in \cite{Wang:2016kzp} is a good approximation for field excursions $\phi \leq \Lambda_h$. Finally, 
our potentials are possibly multi-valued in looping trajectories. We discard such realizations using the procedure in Appendix \ref{sec:potconst}.

As explained in \cite{Fyodorov:2004, Bray:2007tf,Easther:2016ire, Bachlechner:2014rqa} and above,  the correct PDF for a RGF has to incorporate the correlation between the height of the potential $V$ and the Hessian ($\mathcal{H}_{IJ}\equiv \partial^2 V/{\partial \phi^I \partial \phi^J}$.)  For this reason, in this work, we evolve the potential differently to reproduce, in the stationary limit,  the correct  probability density: 
\begin{align}
    \begin{split}
    {\cal P}(V,\mathcal{H}_{IJ}) &\propto e^{-Q} \\
    Q &= (V-\bar{V})^2\frac{N_f+2}{4 V_0^2} + (V-\bar{V})\frac{ \Lambda_h^2}{2 V_0^2} \Tr{\mathcal{H}} + \frac{\Lambda_h^4}{ 4 V_0^2} \Tr{\mathcal{H}^2}
    \end{split}
    \label{eq:RGFpdf}
\end{align} This can be accomplished with the choice:

\begin{align}
    \begin{split}
	\langle \delta v_{IJ} \rangle &= \left( -\frac{ \left(V-\bar{V}\right)}{\sqrt{N_f}\Lambda_v^4}\delta_{IJ} -  v_{IJ} \right) \delta s \\
	\langle \delta v_{IJ}^2 \rangle &= \frac{2 V_0^2}{ N_f \Lambda_v^8} \left(1+\delta_{IJ} \right) \delta s
	\end{split} \label{eq:var}
\end{align} where $\delta s = \sqrt{\delta \tilde{\phi}^I \delta \tilde{\phi}_I}$. In the simulations presented here, $\delta v_{IJ}$ has been chosen using a normal distribution with the average and standard deviation that fit (\ref{eq:var}).

After sufficiently many iterations of this procedure, when $N_f \gg 1$,  the Hessian matrix $\mathcal{H}_{IJ}$ will have eigenvalues distributed in a shifted Wigner semicircle \cite{Bray:2007tf}

\begin{align}
 \rho(\lambda) =\left\{
\begin{array}{cl}
\frac{1}{ 2 \pi N_f} \sqrt{4 N_f - \left(\lambda + \frac{V-\bar{V}}{V_0} \right)^2}  & - 2 \sqrt{N_f} -\frac{V-\bar{V}}{V_0} \leq \lambda \leq 2 \sqrt{N_f }- \frac{V-\bar{V}}{V_0}   \\
 0 & \rm{elsewhere}    \\
\end{array}
\right.
\label{eq:shiftedWigner}
\end{align} where we have introduced the dimensionless eigenvalue $\lambda$ \footnote{Be aware that there are several definitions of $\lambda_I$. In \cite{Freivogel:2016kxc}, for example, $\lambda_I$ stand for the eigenvalues of $v_{IJ}$, hence the difference in the $N_f$-dependence of the endpoints of the allowed interval. Both definitions agree on the $N_f$-dependence of the mass spectrum. }:

$$ m^2 =\frac{V_0}{\Lambda_h^2} \lambda $$ The growth of the range of eigenvalues with $N_f$ is the result of the eigenvalue repulsion characteristic of the DBM evolution. The shift of the center of the distribution depends on the relative value of $(V-\bar{V})/V_0$. In this work we will make no assumptions about the relative sign of $V-\bar{V}$, we will explore both signs. It has been suggested  \cite{Bachlechner:2014rqa} that $V_0$ could be a function of $N_f$, but we have restricted our simulations to an $N_f$-independent $V_0$.

The Fokker-Planck-Smoluchowski equation can be solved exactly for $(V-\bar{V})/V_0$  constant, which is a reasonably good approximation during inflation. It  gives the following evolution of the Hessian in field space for an arbitrary $N_f$:

\begin{align}
    \begin{split}
    {\cal P}(V,\mathcal{H}_{IJ},s) &=\frac{\rm{Const}}{(1-q^2)^{N_f(N_f+1)/2}} e^{-Q (s)} \\
    Q &= \frac{\Lambda_h^4}{ 4 V_0^2 (1-q^2)} \Tr[\mathcal{H}- q \mathcal{H}' + (1-q)\frac{V-\bar{V}}{\Lambda_h^2} \mathbb{1}]^2\end{split}
    \label{eq:RGFTEpdf}
\end{align} where $q=e^{-s}$ and $\mathcal{H}_{IJ}'=\mathcal{H}_{IJ}(s=0)$.
The stationary limit corresponds $q \rightarrow 0$. 
Here, $\Lambda_h$ sets the distance scale over which the Hessian will evolve from arbitrary initial conditions to being approximately in the shifted Gaussian Orthogonal Ensemble (GOE) (\ref{eq:RGFpdf}) \footnote{Note that, despite being a diffusion process, for small $s$ the change in the matrix element PDF is linear in $s$, not a square root.}.
In the simulations we have done, the field excursion rarely exceeds $2 \Lambda_h$ before the end of inflation, thus we never reach this point. 

The mass distribution (\ref{eq:shiftedWigner}) gives an average separation between the masses \cite{Bjorkmo:2017nzd}

\begin{align}
\frac{(m_{\rm{max}}^2-m_{\rm{min}}^2)/N_f}{H^2}=12 \frac{1}{\sqrt{N_f}} \frac{V_0}{V}\left( \frac{M_P}{\Lambda_h} \right)^2
\label{eq:eiggap}
\end{align}
In  our realizations we have assumed parameters roughly of $5\leq N_f \leq 100$, $0.2\Mpl \leq \Lambda_h \leq 0.4 \Mpl$, $V_0 \sim V$ (our full parameter selections are available in Table \ref{tab:data}).
For this reason one might expect a purely single field behavior.  It is important to keep in mind, however,  that the masses fluctuate (\ref{eq:RGFTEpdf}) so in some realizations this separation is much smaller than $H^2$ (cf. Section \ref{sec:singleresults}). 

\FloatBarrier
\subsection{What is the shape of the potential?}\label{sec:PotentialCheck}
\par
\begin{figure}[h]
    \centering
    \includegraphics[width=0.8\textwidth]{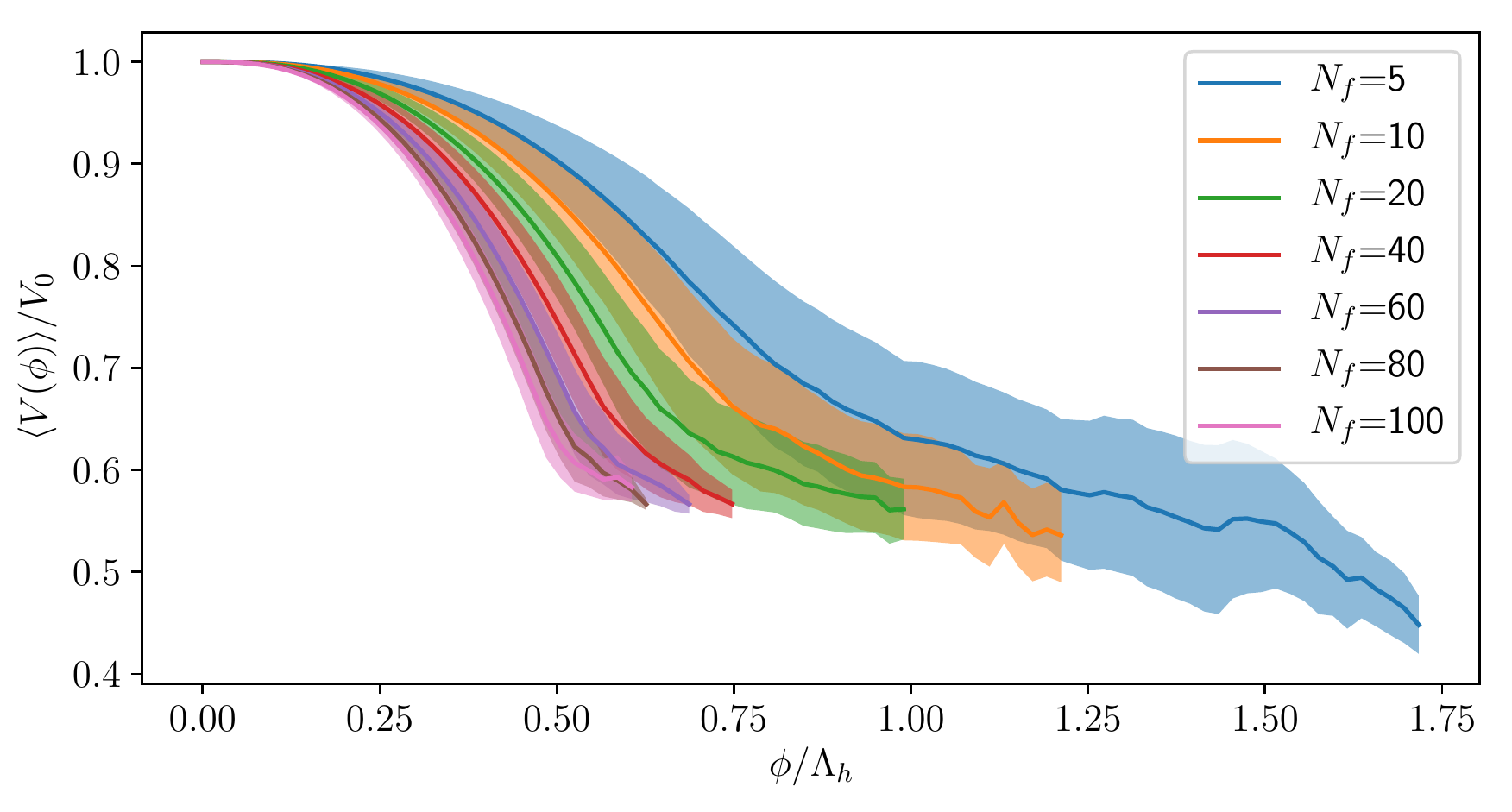}
    \caption{The average shape of the potential along the inflationary trajectory in dataset 1 (see Table \ref{tab:data}). Shaded regions show one standard deviation around the mean value. The excursion becomes shorter and the potential steeper at higher $N_f$.}
    \label{fig:shapeNf}
\end{figure}

\begin{figure}[h]
    \centering
    \includegraphics[width=0.8\textwidth]{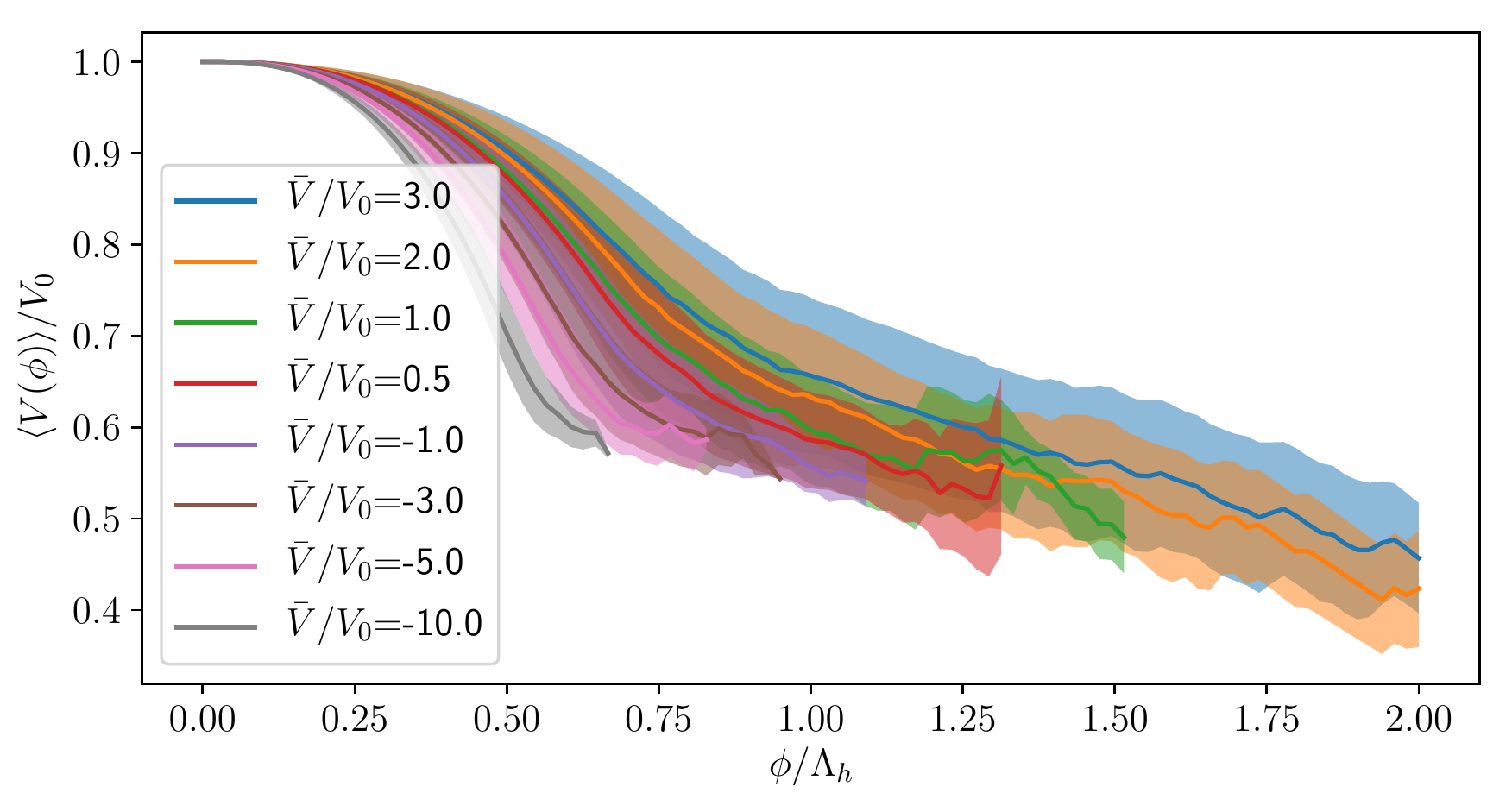}
    \caption{The average shape of the potential along the inflationary trajectory in datasets 4 and 5 (see Table \ref{tab:data}). Shaded regions show one standard deviation around the mean value. The excursion becomes shorter and the potential steeper as the mean becomes more negative. When $\bar{V} > V_0$, the potential shows a markedly slower decay, as a consequence of the eigenvalues being pushed positive (c.f. (\ref{eq:RGFTEpdf})). }
    \label{fig:shapeV}
\end{figure}

 In our construction, we directly prescribe a correlation between the potential and Hessian, precisely (\ref{eq:RGFpdf}).
However a true RGF, described by (\ref{eq:RGFcorr}), has correlations between all even-ordered Taylor coefficients, and all odd-ordered Taylor coefficients.
The odd and even coefficients are uncorrelated \cite{Yamada:2017uzq,Masoumi:2016eag,Masoumi:2017xbe}.
We do not impose any type of correlation on the first and third derivatives.
The potentials in our construction do not have well-defined third derivatives except along the trajectory, and the first derivatives are given by imposing differentiability between patches
\footnote{It may be possible to create DBM potentials with well-defined third derivatives using the procedure of \cite{Battefeld:2014qoa}. However, we do not know how to incorporate the necessary RGF correlations.}. The RGF method and the modified DBM method generate two families of potentials with the same masses at critical points.
It is a natural question to ask, then, how closely our construction matches a RGF. Unfortunately, checking this point is not straightforward. Though, it is possible to compute $\langle V(0) V(\phi)\rangle$ along the classical trajectory, this quantity is not readily comparable to (\ref{eq:RGFcorr}). In our construction, we do not enforce the constraint $\langle V(0)\rangle =\bar{V}$, in fact V(0) varies little, thus $\langle V(0) V(\phi)\rangle$ effectively computes the average shape of the potential. This shape depends on $\bar{V}$ as can be seen in figure \ref{fig:shapeV}. The dependence on $N_f$ is given in figure \ref{fig:shapeNf}.
The potentials along the inflationary trajectory were averaged together over their excursion in field space, and realizations were excluded from the average once the plotted excursion exceeded the excursion travelled during that realization. Points at the extreme right of the plot are an average of only a few realizations, and so are more ragged.
The evolution of the potential in the directions perpendicular to the classical trajectory depends on the mass matrix. We study this evolution in the next section and in figure \ref{fig:single_vabs}.

\FloatBarrier
\subsection{Monte Carlo}
\begin{figure}[h]
    \centering
    \includegraphics[width=\textwidth]{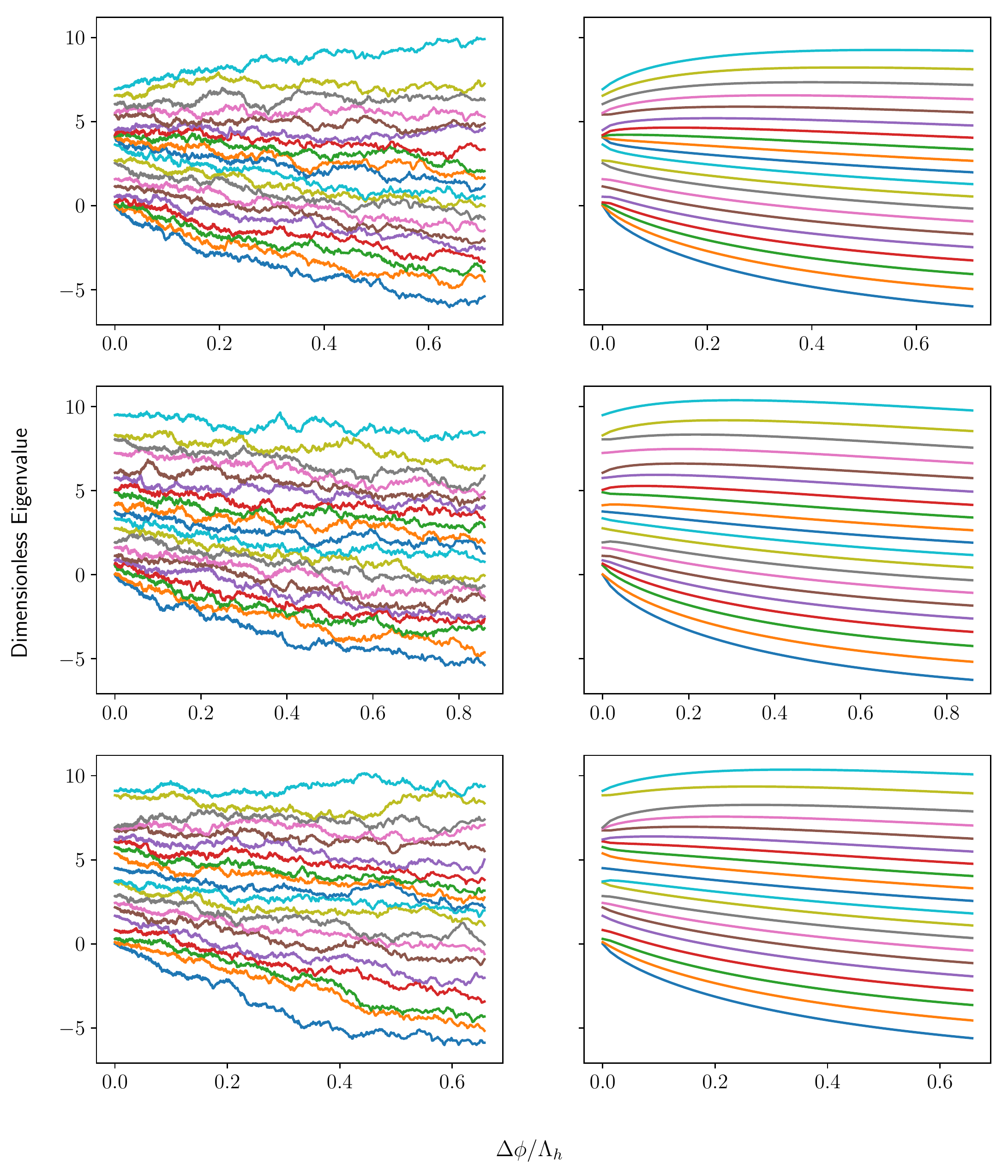}
    \caption{(Left) The eigenvalues of $v_{IJ}$ of three 20-field models drawn from dataset 1, shown versus the covered arclength during inflation. These realizations had excursions of (from top to bottom) $\Delta \phi  =  0.707 \Lambda_h, 0.859 \Lambda_h$ and $0.657 \Lambda_h$.
    (Right) The Monte-Carlo maximization of (\ref{eq:RGFTEeigpdf}), given the same initial eigenvalue spectrum as left, over the same number of correlation lengths.}
    \label{fig:montecarlo}
\end{figure}

As a check on our random walk, we numerically maximized the PDF (\ref{eq:RGFTEpdf}) in Figure \ref{fig:montecarlo}.
We used the \texttt{scipy.optimize.basinhopping} algorithm \cite{doi:10.1021/jp970984n} to maximize the probability distribution.
To reduce the degrees of freedom, we worked with the eigenvalue PDF, rather than the Hessian PDF directly.
The PDF of the eigenvalues can be found by standard techniques once given the Hessian PDF \cite{Mehta2004,Mehta1991}.

We find:
\begin{align}
    \begin{split}
    -\log{{\cal P}(\vec{x},q)} = \frac{\Lambda_h^4}{4 V_0^2 (1-q^2)} &\left( \vec{x}.\vec{x} - 2 q \vec{x}.\vec{x}^\prime + q^2 \vec{x}^\prime.\vec{x}^\prime + 2 (\vec{x}-q\vec{x}^\prime ) (1-q) \frac{V-\bar{V}}{2 \Lambda_h^2} \right) + \\
    &- \sum_{I<J}\,\log{\left| x_I - x_J \right|}+ C(q)
    \end{split}
    \label{eq:RGFTEeigpdf}
\end{align}
where $C(q)$ is an overall time-dependent normalization, and is unimportant for the algorithm.
To avoid numerical underflow/overflow issues we maximized the logarithm of the PDF rather than the PDF itself. The logarithm is monotonically increasing over the positive reals, so these results are equivalent.

Unfortunately, (\ref{eq:RGFTEpdf}) and (\ref{eq:RGFTEeigpdf}) are only exact in the limit of a constant potential.
However, they provide a good approximation -- few realizations change the potential by more than $50\%$ during their evolution, and a change of several times the initial value is necessary to substantially shift the distribution (cf. (\ref{eq:shiftedWigner})).
The realizations shown in Figure \ref{fig:montecarlo} have $V(t_\text{end}) / V(t_0) = 0.56,0.53,0.57$ from top to bottom.
The potential was assumed to be constant and equal to its initial value during numerical maximization.

The agreement is qualitative.
In any individual realization, there is a substantial amount of variance in the individual eigenvalue paths away from their most likely locations.
As we show in section \ref{sec:perturbations}, it is precisely this variance that creates a small fraction of realizations with high isocurvature power.

\FloatBarrier
\section{Background Evolution}

\subsection{Conditions for Inflation}

Ideally, we could generate many globally-defined realizations of the landscape, sample points with high tunneling probability, attempt inflation at all of them, and observe the distributions of observables from trajectories with $>60$ efolds.
There are several obstacles to this procedure in the construction in section \ref{sec:potentials}.
\begin{itemize}
    \item they are ill-defined globally, so cannot be blindly sampled.
    \item they have an unstable oscillatory growth beyond a correlation length.
    \item an arbitrary point in an RGF potential with zero mean is unlikely to generate $60$ efolds of inflation.
\end{itemize}
Because of the above difficulties, we require selecting initial conditions for inflation by hand, rather than sampling the landscape.

We compute the slow roll parameters in terms of the potential as 
\begin{align}
\epsilon_V &\equiv \frac{\Mpl^2}{2} \frac{V_{,I} V_{,I}}{V^2} \\
\eta_V &\equiv \Mpl^2 \frac{\mathrm{Min(Eig(}V_{IJ}\mathrm{))}}{V} 
\end{align}

The shape of the potential around the initial inflationary point is not known a priori. 
With a sufficiently high potential, $\epsilon_V$ and $\eta_V$ can always be made sufficiently small to make inflation last any desired length of time.
According to previous work \cite{Marsh:2013qca,Freivogel:2016kxc,Dias:2016slx,Bjorkmo:2017nzd}, we expect inflation to occur most often at saddle points in a RGF.
For this reason, in this work we exclusively study saddle-point inflation.
In terms of the initial potential parameters this translates to $v_I$ small and random in direction, and the eigenvalues of $v_{IJ}$ being all positive with the smallest approximately zero.

The most suitable arrangement of masses for prolonged saddle-point inflation is with one massless or slightly tachyonic field, and the others massive.
Such distributions of masses are uncommon in RGF saddle points.
The average mass distribution at a saddle point is the shifted Wigner semicircle (\ref{eq:shiftedWigner}). To get the desired arrangement of masses from (\ref{eq:shiftedWigner}) there are two options. Either inflation happens for a value of  the potential 
 $V \leq \bar{V}-V_0 2 \sqrt{N_f}$  \cite{Bachlechner:2014rqa} or inflation starts at a special point resulting from an upward fluctuation of the masses. 
Recently, Vilenkin and Yamada \cite{Yamada:2017uzq} studied the eigenvalue distributions at saddle points of a RGF, giving a result for the eigenvalue distribution assuming none are negative:
\begin{align}
    \begin{split}
        \xi &=  -\sqrt{2} + 2 \sqrt{1-\frac{N_f}{N_f+2}} \\
        L(\xi) &= \frac{2}{3}\left( \sqrt{\xi^2 + 6} - \xi\right) \\
        P_\mathrm{init}(\mu,\xi) &= \frac{1}{2 \pi} \left( \sqrt{\frac{L(\xi) - \mu}{\mu}} \left( L(\xi) + 2 \mu + 2 \xi \right) \right)
    \end{split}
    \label{eq:eigICs}
\end{align}
where $\mu$ is the normalized dimensionless eigenvalue $\lambda/\sqrt{N_f}$.
\begin{figure}[h]
    \centering
    \includegraphics[width=0.8\textwidth]{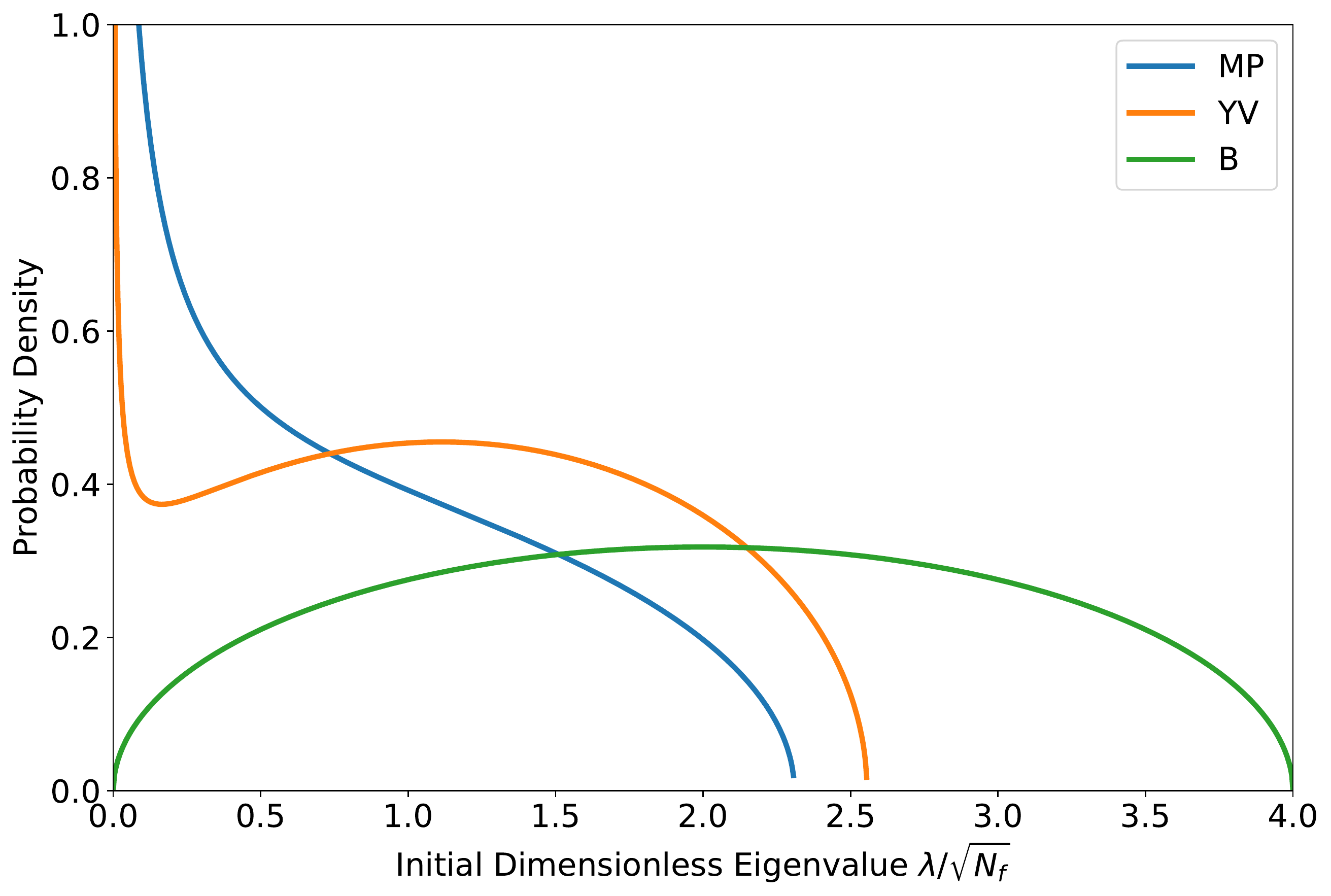}
    \caption{The distributions proposed in \cite{Marsh:2013qca,Yamada:2017uzq,Bachlechner:2014rqa} (MP,YV,and B respectively).}
    \label{fig:mp_dist}
\end{figure}
We chose (\ref{eq:eigICs}) as our initial eigenvalue distribution. This is constant width in $\mu$, but grows in $\lambda$ with $N_f$.

In order to prescribe the initial $\eta_V$, we shifted each set of eigenvalues drawn from this distribution slightly towards the negative, so that the smallest gave the correct value of $\eta_V$.

We prescribe $\epsilon_V$ by hand, ranging from $10^{-12}$ to $10^{-9}$. Larger values of $\epsilon_V$ fail to generate enough e-foldings. For example, for $\epsilon_{V}=10^{-8}$ $91\%$ of the runs had fewer than $55$ efoldings.\footnote{If the claim made in \cite{Obied:2018sgi,Agrawal:2018own} proves to be correct, the range of values for $\epsilon_V$ that we have considered would be inconsistent with our motivation to study multi-field inflation based on a complete UV theory.}
We choose $v_I$ (cf. (\ref{eq:potential})) to be random in direction (uniform over the sphere), with a magnitude chosen to give the correct initial $\epsilon_V$.

The definition of our potential has a degeneracy in the scale of the Taylor coefficients and $\Lambda_V$ \cite{Freivogel:2016kxc,Dias:2017gva}.
We can freely upscale $\Lambda_V^4$, so long as $v_0$, $v_I$,and $v_{IJ}$ are scaled down by the same factor. This leaves the height of the potential and the background equations of motion invariant.
We exploit this degeneracy to tune $\Lambda_V$ post-hoc to give the Planck value of the amplitude of the adiabatic power spectrum while computing perturbations.
We also use it to set the height of the potential with $\Lambda_{V\rm{ init}}$, and set $v_0=1$.
For a fixed size of dimensionless parameters $v_0,v_I$, and $v_{IJ}$, the initial value of $\Lambda_V$,$\Lambda_{V\rm{ init}}$ sets the scale of inflation $H(t=0)$.
Planck constrains the average Hubble parameter $H$ to be no larger than $10^{-4} \Mpl$.
Motivated by this, we choose an initial $\Lambda_{V\rm{ init}} = \mathcal{O}(10^{-1.8}\Mpl)$ (see Table \ref{tab:data}). 

\subsection{Evolution}
\label{sec:evolution}

We numerically solve the coupled equations of motion in cosmic time:
\begin{align}
	\begin{split}
	    H^2 = \frac{1}{3 M_p^2} \left(\frac{\dot{\phi}^I \dot{\phi}_I}{2} + V(\vec{\phi}) \right) \\
	    \ddot{\phi}^I + 3 H \dot{\phi}^I + V^{,I} = 0.
    \end{split}
    \label{eq:backgroundEoM}
\end{align}

Here $\partial_I V \equiv V_{,I}$. 
We choose inflation to start in slow-roll, with $\ddot{\phi}^I = 0$ at $t=0$, and the initial velocity $\dot{\phi}^I = \frac{-V^{,I}}{3 H}$ at $t=0$.

Inflation occurs as long as 
\begin{align}
\epsilon_H \equiv -\frac{\dot{H}}{H^2} < 1.
\end{align}
We evolve (\ref{eq:backgroundEoM}) until $\epsilon_H = 1$ or a numerical time limit is reached, at which point we perform a few tests. If any of these fail, we save the background evolution but do not solve the perturbations' equations of motion.
\begin{itemize}
    \item The time limit was not reached, and $\epsilon_H(N_\textrm{end})=1$.
    \item We cover at least 5 efolds before imposing the earliest mode in Bunch-Davies.
    \item The trajectory has no self-intersections.
    \item The differential equation solver gave no errors during the evolution.
\end{itemize}
\subsection{Eigenvalue Gap}
Bjorkmo and Marsh \cite{Bjorkmo:2017nzd} observed a distinct gap between the two lowest effective masses in their RGF potential construction. This gap is noticeably absent from our construction, and from our Monte Carlo maximization of the Hessian's PDF in figure \ref{fig:montecarlo}.
Following the argument of Bjorkmo and Marsh, this is expected: our potentials do not include a well-defined third derivative term, and it is precisely the structure of this term which gives rise to the eigenvalue gap in their construction.

Vilenkin and Yamada \cite{Yamada:2017uzq} predicted the eigenvalue gap in a VY distribution to go like $1/\sqrt{N_f}$, by taking the width of the distribution to go like $\sqrt{N_f}$ and dividing by $N_f$ for the $N_f$ eigenvalues.
\begin{figure}[h]
    \centering
    \includegraphics[width=0.9\textwidth]{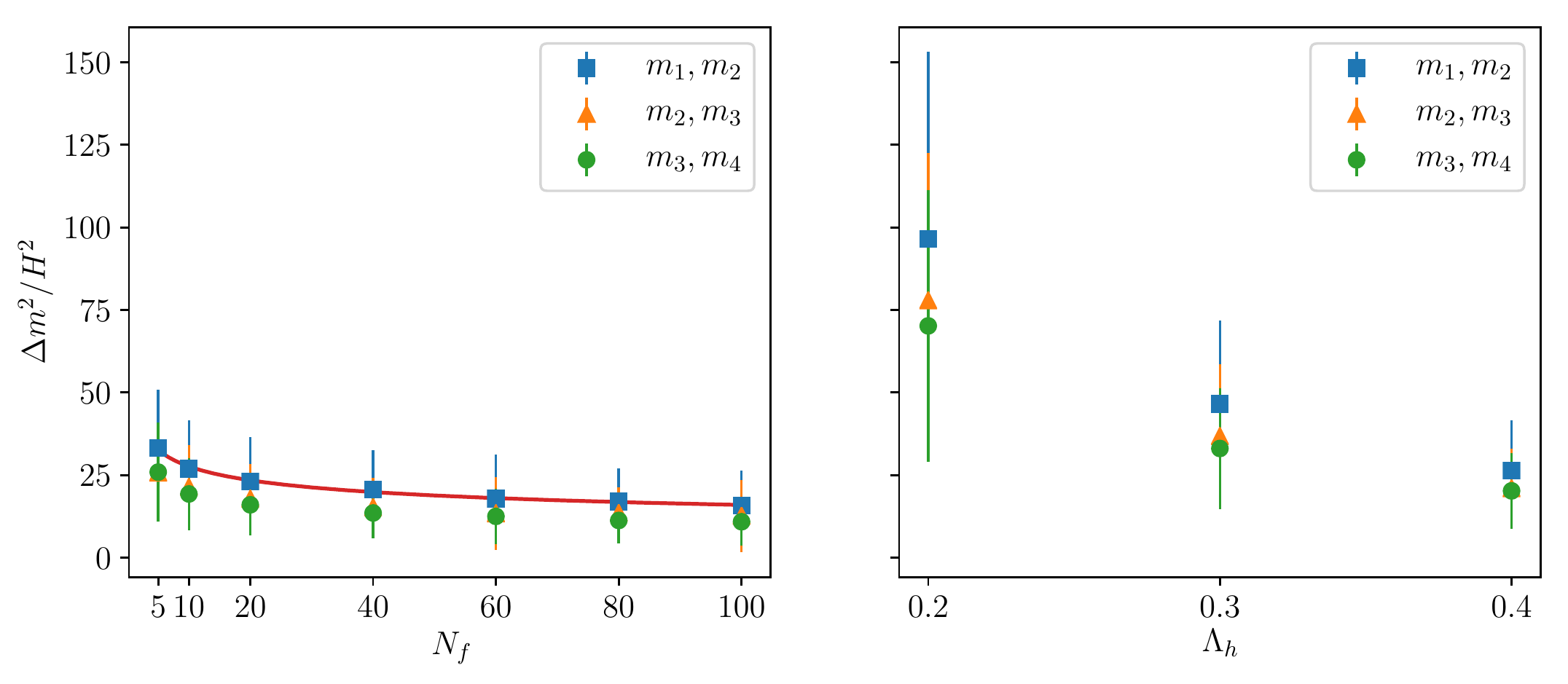}
    \caption{The gap between the lowest eigenvalue pair, the next lowest, and the next next lowest, at the end of inflation. We show dataset 1 to the left and dataset 6 to the right. As a function of $N_f$, the gap in physical units shrinks approximately as $N_f^\delta$, with $\delta = -0.24$.}
    \label{fig:eiggap}
\end{figure}

However the width of the distribution (c.f. (\ref{eq:eigICs})) goes like $L(\xi)\sqrt{N_f}$, which only approximates $\sqrt{N_f}$ in the large-$N_f$ limit. $L(\xi)$ varies significantly over the range $5\leq N_f\leq 100$. A power law fit of $N_f^{0.73}$ is a decent approximation over this range of $N_f$.
We expect the initial eigenvalue gap to go like the width divided by $N_f$, or approximately $N_f^{-0.27}$.
In the equilibrium distribution, we expect the standard Wigner eigenvalue gap, proportional to $1/\sqrt{N_f}$ (\ref{eq:eiggap}).
However, on average we cover too few correlation lengths to reach equilibrium, and we expect a scaling much more like the initial distribution than the final one.
We observe approximately $N_f^{-0.27}$ scaling in figure \ref{fig:eiggap}.

\FloatBarrier
\section{Perturbations}
\label{sec:perturbations}
Inflationary observables are given by perturbations around the classical trajectory.

Multi-field inflation is fundamentally different than single-field inflation.
As shown in  \cite{GrootNibbelink:2000vx,GrootNibbelink:2001qt,Schutz:2013fua}, scalar perturbations orthogonal to the inflationary trajectory can become excited, and source super-horizon evolution of the adiabatic mode.
These so-called isocurvature modes can source the observable scalar power spectra at different rates, and lead to a CMB power spectrum different than observations.
The bound from Planck is that all isocurvature power is at most $\sim 10^{-2}$ of the adiabatic power \cite{Ade:2015lrj}.

An additional complication comes from reheating.
During (p)reheating, isocurvature modes can decay or source additional adiabatic evolution.
In some scenarios, primordial isocurvature power can be preserved by reheating, or amplified by it, and lead to observable consequences \cite{Hotinli:2017vhx}.

As we describe in section \ref{sec:data}, our model allows significant isocurvature power at the end of inflation, and could potentially lead to observable consequences.
Large isocurvature power is rare, but not atypical.
This result is different from \cite{Bjorkmo:2017nzd,Dias:2017gva}, in which isocurvature can always decay to unobservable levels in RGF and DBM models.

\subsection{What generates isocurvature?}
It is helpful to analyze the perturbations in a natural basis \cite{GrootNibbelink:2000vx, GrootNibbelink:2001qt, Baumann:2014nda, Schutz:2013fua}

$$ \phi^{I}_{(n)} \equiv \partial_t^{(n)}\phi^{I} \hspace{3ex} \rm{for} \hspace{3ex} 1 \leq n \leq N_f \label{dynbasis}$$  From this basis, we can construct an orthonormal basis $\{e^I_{(1)}, \cdots,e^I_{(N_f)} \}$ using the Gram-Schmidt process. The first vector $e^I_{(1)}$ is tangent to the classical inflaton trajectory while the second $e^I_{(2)}$ is parallel to the transverse acceleration. 

\begin{eqnarray}
	e^I_{(1)} &\equiv& \hat{\sigma}^I = \frac{\dot{\phi^I}}{\dot{\sigma}}, \hspace{3ex} \dot{\sigma}^2 \equiv \sum_I (\dot{\phi^I})^2 \\
	e^I_{(2)} &\equiv &\hat{s}^I= \frac{\omega^I}{\omega},  \hspace{3ex} \omega^I=\dot{\hat{\sigma}}^I \hspace{3ex} \omega^2 \equiv \sum_I (\omega^I)^2 \label{eq:defOmega}
\end{eqnarray}  The field fluctuations along these directions  

$$ Q_{\sigma}= e_{(1)_I} Q^I, \hspace{3ex}   Q_s=e_{(2)_I} Q^I$$ play important roles. $Q_{\sigma}$ is the adiabatic perturbation and a measure of the comoving curvature perturbation
\begin{align}
{\cal R}= \frac{H}{\dot{\sigma}} Q_{\sigma} \label{eq:RQsigma}
\end{align}
Its equation of motion is:
\begin{eqnarray*}
	\ddot{Q}_{\sigma} &+& 3 H \dot{Q}_{\sigma}+ \left[  \frac{k^2}{a^2} + {\cal M}_{\sigma \sigma} - \omega^2 -\frac{1}{M_P^2 a^3}    \frac{d}{dt} \left( \frac{a^3 \dot{\sigma}^2} {H}\right)        \right] Q_{\sigma}  \\ \\
	&=&\frac{d}{dt}(\omega Q_s)-2 \left(  \frac{V_{, \sigma}}{\dot{\sigma}} + \frac{\dot{H}}{H}\right) \omega Q_s
\end{eqnarray*}
while that of $Q_s$ is
$$\ddot{Q}_{s} + 3 H \dot{Q}_{s}+ \left[  \frac{k^2}{a^2} + {\cal M}_{ss} + 3 \omega^2       \right] Q_s=4 M_P^2 \frac{\omega}{\dot{\sigma}} \frac{k^2}{a^2} \Psi + \cdots $$
where $\Psi$ is the metric perturbation in the Newtonian gauge and the ellipsis stands for the couplings to the remaining isocurvature modes.
We will refer to $Q_s$ as the first isocurvature mode, and all others as higher isocurvature modes.
The masses are projections of the mass matrix defined in (\ref{eq:massmatrix}).

\begin{eqnarray*}
	{\cal M}_{\sigma \sigma} &=& \sigma^I \sigma^J M_{IJ} \\
	{\cal M}_{s s} &=& s^I s^J M_{IJ} \\
\end{eqnarray*}
By analogy with (\ref{eq:RQsigma}) we define
\begin{align}
{\cal S}= \frac{H}{\dot{\sigma}} Q_s \label{eq:SQsigma}
\end{align}
The time evolution of ${\cal R}$  and ${\cal S}$, in the limit $k^{2}/(aH)^2 \ll 1$, is
\begin{eqnarray}
\dot{{\cal R}} &\simeq& \alpha H {\cal S}  \hspace{9ex} \alpha=\frac{2 \omega}{H} \nonumber \\ \\
\dot{{\cal S}} & \simeq& \beta H {\cal S} + \cdots  \hspace{3ex} \beta= - 2 \epsilon_H -\frac{M_P^2{\cal M}_{ss}}{ V} + \frac{M_P^2{\cal M}_{\sigma \sigma}}{ V}-\frac{4 \omega^2}{3 H^2} \nonumber \label{eq:RSevol}
\end{eqnarray}
while $\alpha$ is always positive, the sign of $\beta$ is uncertain and may change as a function of time.
The evolution of ${\cal S}$, depends  on the sign of $\beta$ and on the terms described here by the ellipsis.
During the background evolution we compute and save $\beta$ and $\omega$, as well as the closely related quantity
\begin{align}
\eta_\perp \equiv -V_{,I} e_{(2)}^I / (H \dot{\sigma}) \propto \omega^2 / H^2.\label{eq:etaperp}
\end{align}
Our simulations show that in a small percentage of cases (see figure \ref{fig:iso_percent}) the isocurvature modes grow outside the horizon, generating a power that can be as high as, or higher than,  the Planck upper limit of $10^{-2}$ times the adiabatic power.
Even if this result is not in conflict with the  bounds from Planck  on the isocurvature power, it should cast doubts on our ability to make  observable  predictions  independently of the reheating model.
Isocurvature modes can source non-gaussianities.
In this work we have not checked if the amount of non-gaussianity generated by these modes is already excluded because DBM is not suited for this computation.
In particular, the third derivatives of our potentials are ill-defined.
Other methods, such as \cite{Bjorkmo:2018txh}, are better suited for this task.

\subsection{The transport method}
\par We compute the power spectrum of perturbations around the classical inflationary trajectory through the transport method \cite{Dias:2015rca,Dias:2016slx,Dias:2017gva}.
This method is generally numerically efficient and stable.
Our potentials are not known analytically and the power spectrum of scalar perturbations is not guaranteed to have a simple form, so we need to compute $P_\zeta(k)$ for a wide range of $k$s, each requiring the evolution of $\mathcal{O}(N_f^2)$ ODEs.
Our implementation of this method, once optimized, was able to compute perturbations for $N_f=100$ models adequately quickly \footnote{On a \texttt{stampede2} SKX compute node, computing perturbations for a $N_f=100$ model took $\sim$ 1 core-hour.}. 

In this section lowercase Latin indices $a,b,\ldots$ will run from $0,\ldots,2N_f-1$, while uppercase Latin indices will be consistent with the rest of this paper, and run from $0,\ldots,N_f-1$.
Occasionally we will split a lowercase index into an uppercase index and its bar.

We write the perturbations (in spatially-flat gauge) as 
\begin{align}
    \phi^I(t) = \phi^I(t)_\text{classical} + Q^I (t).
\end{align}
We can expand the inflationary action to quadratic order in the perturbations:
\begin{align}
    S_{(2)} &= \frac{1}{2} \int \frac{\text{d}^3 k}{(2\pi)^3} \text{d}t \, a^3 \left( \delta_{IJ} \partial_t Q^I (\pmb{k}) \partial_t Q^J(\pmb{-k}) - \left[ \frac{k^2}{a^2}\delta_{IJ} + M_{IJ} \right] Q^I (\pmb{k}) Q^J(\pmb{-k}) \right)
\end{align}
where the mass matrix, $M_{IJ}$, is  
\begin{align}
    M_{IJ} \equiv V_{,IJ} - \frac{1}{a^3 \Mpl^2} \partial_t \left( a^3 \frac{\dot{\phi_I}\dot{\phi_J}}{H} \right)
    \label{eq:massmatrix}
\end{align}
The field perturbations have non-canonical conjugate momenta
\begin{align}
    \delta \pi^I &\equiv \partial_N Q^I
\end{align}
where $\partial_N$ is a derivative with respect to the number of efolds ($N$) since the start of inflation.
For notational convenience, we define the concatenation of the field and momenta perturbations $X \equiv \{ Q ,\delta \pi \} $.
The equations of motion for the perturbations can be written, to tree level, as \cite{Dias:2015rca,Dias:2017gva}
\begin{align}
    \partial_N X^a = u^a_b X^b + \ldots
\end{align}
where 
\begin{align}
    u^a_b &= \begin{pmatrix}
    0 & \delta^{A}_{\bar{B}} \\
    -\delta^{\bar{A}}_{B} \frac{k^2}{a^2 H^2} - \frac{M^{\bar{A}}_{B}}{H^2} & (\epsilon_H - 3) \delta^{\bar{A}}_{\bar{B}}
    \end{pmatrix}.
\end{align}

In the transport method, we evolve the two-point correlator of the perturbations directly, rather than solving for the modes themselves.
This is possible by applying the chain rule: 
\begin{align}
\partial_N \langle X^a X^b \rangle &= \langle (\partial_N X^a) X^b \rangle + \langle X^a (\partial_N X^b) \rangle
\label{eq:ehrenfest}
\end{align}
If we define the two-point correlator as
\begin{align}
    \langle X^a X^b \rangle &= (2\pi)^3 \delta(\vec{k}+\vec{k}^\prime) \frac{\Sigma^{ab}}{k^3},
\end{align}
its equation of motion reads
\begin{align}
    \partial_N \Sigma^{ab} &= u^a_c \Sigma^{cb} + u^b_c \Sigma^{ac} + \ldots
    \label{eq:Sig_transport}
\end{align}
In practice, we do not solve (\ref{eq:Sig_transport}) directly, rather we solve for the propagator $\Gamma^a_b(N,N_0)$
\begin{align}
    \frac{d \Gamma^a_b }{dN} &= u^a_c \Gamma^c_b\\
    \Sigma^{ab}(N) &= \Gamma^a_c (N,N_0) \Gamma^b_d (N,N_0) \Sigma^{cd}(N_0)
\end{align}
where $\Gamma^a_b(N,N_0)$ propagates the dimensionless two-point function $\Sigma^{bd}$ from a time with known initial conditions, $N_0$, to any later time, $N$.

The initial conditions for $\Sigma^{ab}$ depend on the mode.
When a mode is sufficiently subhorizon, $\Sigma^{ab}$ will be approximately independent of any spatial curvature, and can described as approximately Bunch-Davies.
For a mode with wavenumber $k$, this is given by
\begin{align}
    \Sigma^{ab}|_\text{BD} &= 
    \begin{pmatrix}
    \frac{H^2 \delta^{IJ}}{2} \left| k \tau \right|^2 & -\frac{H^2 \delta^{\bar{I}J}}{2} \left| k \tau \right|^2 \\
    -\frac{H^2 \delta^{I\bar{J}}}{2} \left| k \tau \right|^2 & \frac{H^2 \delta^{\bar{I}\bar{J}}}{2} \left| k \tau \right|^4
    \end{pmatrix}.
\end{align}
We impose these initial conditions $5$ efolds before the mode reaches horizon exit $k=aH$.
In numerical tests, this was found to be sufficient to leave the power spectrum invariant.

Note that these initial conditions are only valid at a time of approximate slow-roll \footnote{see discussion around (3.9) of \cite{Dias:2015rca}}.This is satisfied in all of our realizations ($\eta_\perp \sim 0.01$ at this time in most realizations, maximum $\sim 0.1$). As an additional check, the $\eta_\perp$ at horizon exit of the pivot-scale mode is shown versus isocurvature in figure \ref{fig:corr_matrix} -- there is no correlation.

\subsection{Observables}
In order to compute physical, gauge-invariant quantities, we need a gauge transformation out of spatially flat gauge.
One common scalar observable is $\zeta$, the adiabatic scalar perturbation on surfaces of constant density.
The relevant transformation from spatially-flat gauge into constant-curvature gauge can be written as \footnote{There exist several equivalent forms of this gauge transformation. One other, used by \cite{Dias:2015rca,Dias:2017gva} is $$ N_a = \{\frac{1}{2 \epsilon_H} \frac{V_{,A}}{V}, \frac{1}{2\epsilon_H(3-\epsilon_H)} \frac{\dot{\phi}^{\bar{A}}}{H \Mpl^2}\}.$$ We use (\ref{eq:gaugetransform}) because it has zeros in the momenta half of the vector, and this allowed us to compute an orthogonal space for the isocurvature power spectrum.}
\begin{align}
    N_a = \{-\frac{1}{2\epsilon_H} \frac{\dot{\phi}^A}{H}, 0\}.
    \label{eq:gaugetransform}
\end{align}
The adiabatic spectrum, for any mode $k$ at any time $N$, is then given by
\begin{align}
    P_\zeta(k,N) = \frac{1}{2\pi^2} N_a(N) N_b(N) \Gamma^{a}_c(N,N_0) \Gamma^b_d(N,N_0) \Sigma^{cd}(k,N_0)
\end{align} Note the factor of $\frac{1}{2\pi^2}$, which we include for agreement with Planck. This factor is notably absent from previous work using the transport method \cite{Dias:2015rca,Dias:2016slx,Dias:2017gva}.
We fit the power spectrum in $k$-space to measure the spectral index and its running 
\begin{align}
    P_\zeta(k,N_\text{end}) = A_\zeta \left( \frac{k}{k_\star} \right)^{n_s - 1 + \frac{\alpha_s}{2} \log{k/k_\star} }
\end{align}

\subsubsection{Isocurvature Perturbations}

One other relevant observable in multi-field models of inflation is the isocurvature power, or the scalar power in directions orthogonal to the fields' velocity.
We first construct an orthonormal basis for the tangent space to the fields' velocity, with basis vectors $v^\alpha_I$, where $\alpha$ labels the $N_f-1$ tangential directions, and $I$ labels the $N_f$ components of each basis vector.
We then trace over the field-field quadrant of the 2-point function matrix $\Sigma^{IJ}(N)$.
\begin{align}
    P_{\text{iso}} \equiv \frac{1}{2\pi^2} \left( \frac{H}{\dot{\sigma}} \right)^2 \delta_{\alpha\beta} v^\alpha_I(N) v^\beta_J(N) \Gamma^{I}_a(N,N_0) \Gamma^J_b(N,N_0) \Sigma^{ab}(k,N_0) 
    \label{eq:isocurvature_power}
\end{align}
where $\dot{\sigma}^2 \equiv \dot{\phi}^I \dot{\phi}_I$.
The equivalent of the gauge transformation (\ref{eq:gaugetransform}) here are the factors of $H/\dot{\sigma}$.

Note that $P_\text{iso}$, as defined above, is a sum of the power spectra in every direction orthogonal to the adiabatic spectrum.
This includes the first isocurvature mode defined in (\ref{eq:RSevol}), and all higher isocurvature modes.
To our knowledge, an explicit expression for the isocurvature power in the full transport method has not been previously stated in the literature.

\section{Results}
\par Our potential construction has a vast parameter space. In the language of \cite{Bjorkmo:2017nzd}, our hyperparameters were the number of fields $N_f$, the initial values of $\epsilon_V$ and $\eta_V$, and the initial eigenvalue distribution.
In addition there are the RGF parameters, $\bar{V}$ and $\Lambda_h$ (we fix $V_0$ and $\Lambda_V$).
Rather than attempt to sample this vast space uniformly, we sliced it in the following datasets:
\begin{table}[h]
	\centering
	\scalebox{0.8}{
	\begin{tabular}{l|l|l|l|l|l|l}
		dataset \# & $N_f$ & $-\log_{10}{(\epsilon_V|_0)}$ & $\Lambda_h/\Mpl$ & $-\log_{10}{(-\eta_V|_0)}$ & $\bar{V}/V_0 $  & $-\log_{10}{(\Lambda_{V\mathrm{ init}})}$ \\ \hline
		1 & 5,10,20,40,60,80,100 & 12 & 0.4 & 5 & 0 & $1.8$  \\ 
		2 & 10 & 12,11,10,9,8*,7* & 0.4 & 5 & 0 & $1.8$ \\ 
		3 & 10 & 12 & 0.4 & 6,5,4,3,2,1* & 0 & $1.8$  \\
		4 & 10 & 11 & 0.4 & 4 & -1,0,0.5,1,2 & $1.8$  \\ 
		5 & 10 & 12 & 0.4 & 5 & -10,-5,-3,3,5*,10* & $2.0$  \\ 
		6 & 10,40,80 & 12 & 0.1*,0.2,0.3,0.4 & 5 & 0 & $1.8$*, $1.5$  \\
	\end{tabular}}
	\caption{All datasets used in this work. Each value of hyperparameters was run until we had 1000 realizations which computed observables. Parameter combinations with a * were attempted, but successfully computed observables too infrequently to get 1000 usable realizations within computing time constraints. See the discussion in section \ref{sec:evolution} for the conditions under which a realization computed observables.}
	\label{tab:data}
\end{table}
\FloatBarrier
\subsection{A single realization}
\label{sec:singleresults}
To get some intuition for these models, it will be helpful to go through some single realizations in detail.
We consider a typical 80-field model, and an atypical one with large isocurvature.
Both models have the parameters from dataset 1, i.e., an initial $\epsilon_V = 10^{-12}$, an initial eta of $\eta_V = -10^{-5}$, $\Lambda_h = 0.4 M_\text{pl}$, a zero mean, the first derivative of the potential random in direction, and the initial field configuration is in slow roll, with $\ddot{\phi}^I=0$ and $\dot{\phi}^I = - \frac{V_{,I}}{3 H}$.
\begin{figure}[h]
    \centering
    \includegraphics[width=0.9\textwidth]{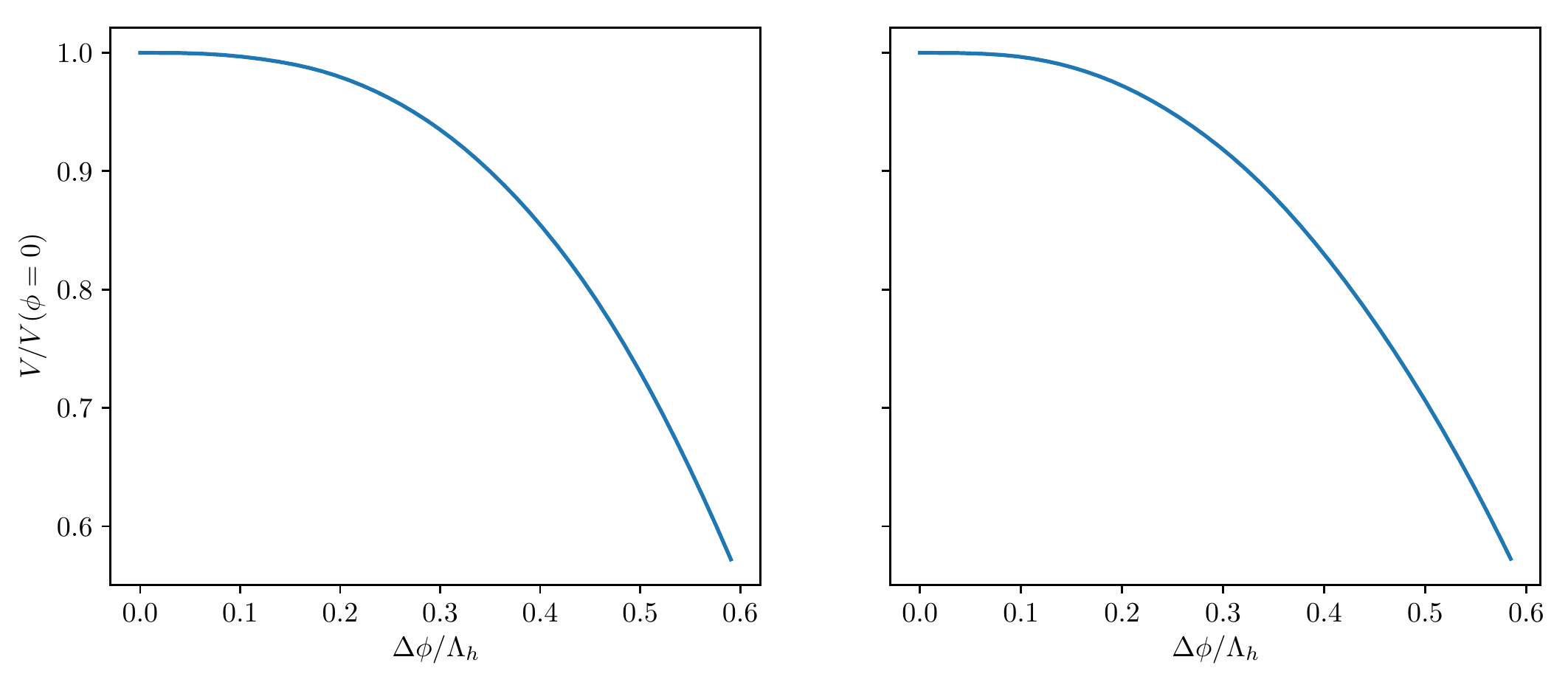}
    \caption{The potential for the two realizations considered here along the trajectory, plotted against arclength, for both the typical (left) and atypical (right) realizations.}
    \label{fig:single_vs}
\end{figure}

The relative change of the potential of the two realizations is presented in Figure \ref{fig:single_vs}.
Despite the patchwork construction of the potential, it is smooth (up to $C^2$) and featureless.
\begin{figure}[H]
    \centering
    \includegraphics[width=0.9\textwidth]{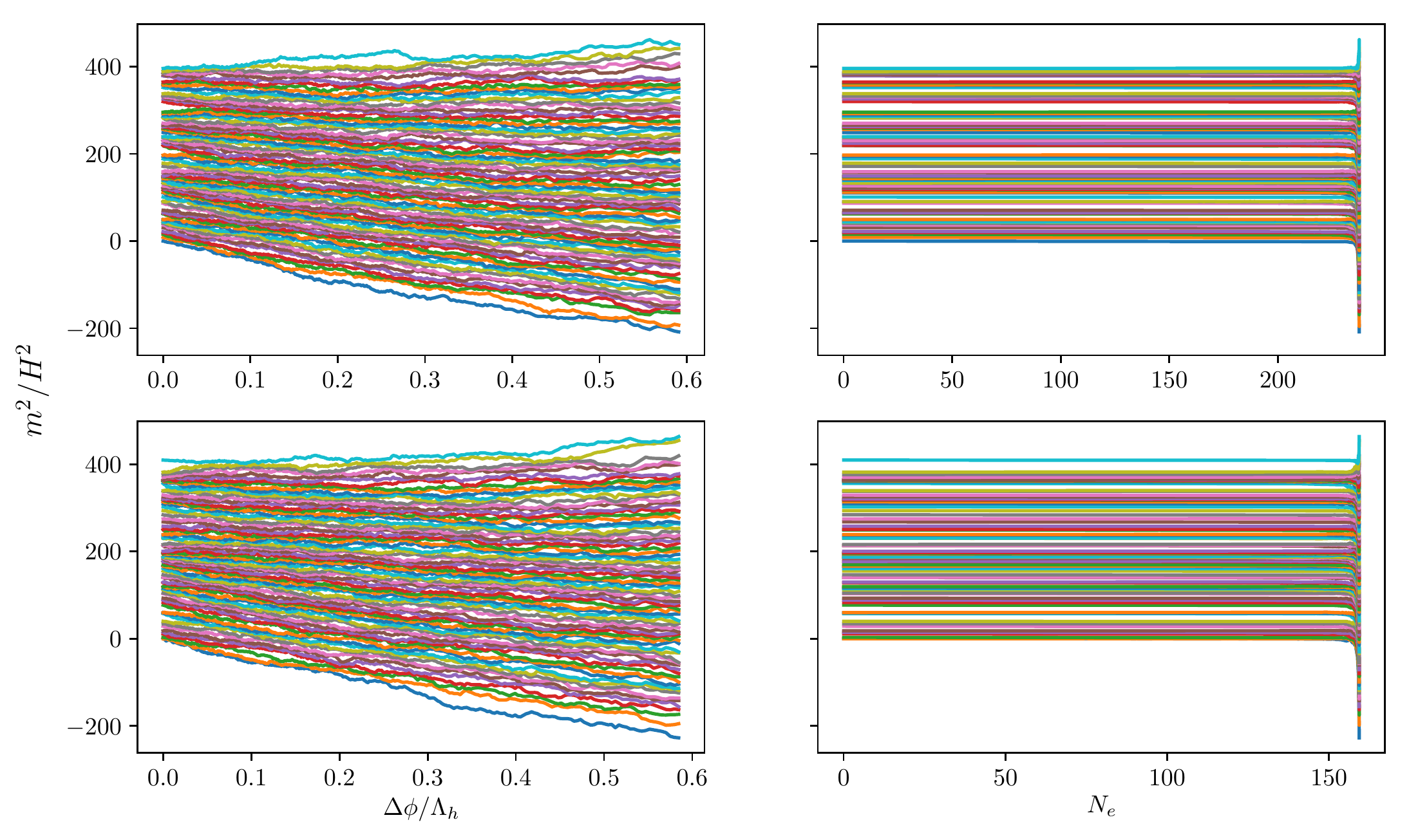}
    \caption{The eigenvalue evolution for the two realizations. Plotted against arclength (left) and efolds (right), for the typical realization (above) and atypical (below). Initial configurations were drawn from (\ref{eq:eigICs}).}
    \label{fig:single_vabs}
\end{figure}

The eigenvalue evolution is presented in Figure \ref{fig:single_vabs}.
The initial eigenvalues, drawn from the VY distribution (see Figure \ref{fig:mp_dist}), evolve via Dyson Brownian motion.
If inflation covered a few $\Lambda_h$, we would expect the eigenvalues to reach a shifted Wigner semicircle (\ref{eq:shiftedWigner}).
However, few realizations have an excursion further than a single correlation length.
These two realizations had excursions of $0.591 \Lambda_h$ and $0.585 \Lambda_h$ respectively.
The field displacement grows slowly at first, then super-exponentially near the end of inflation.

\begin{figure}[H]
    \centering
    \includegraphics[width=0.9\textwidth]{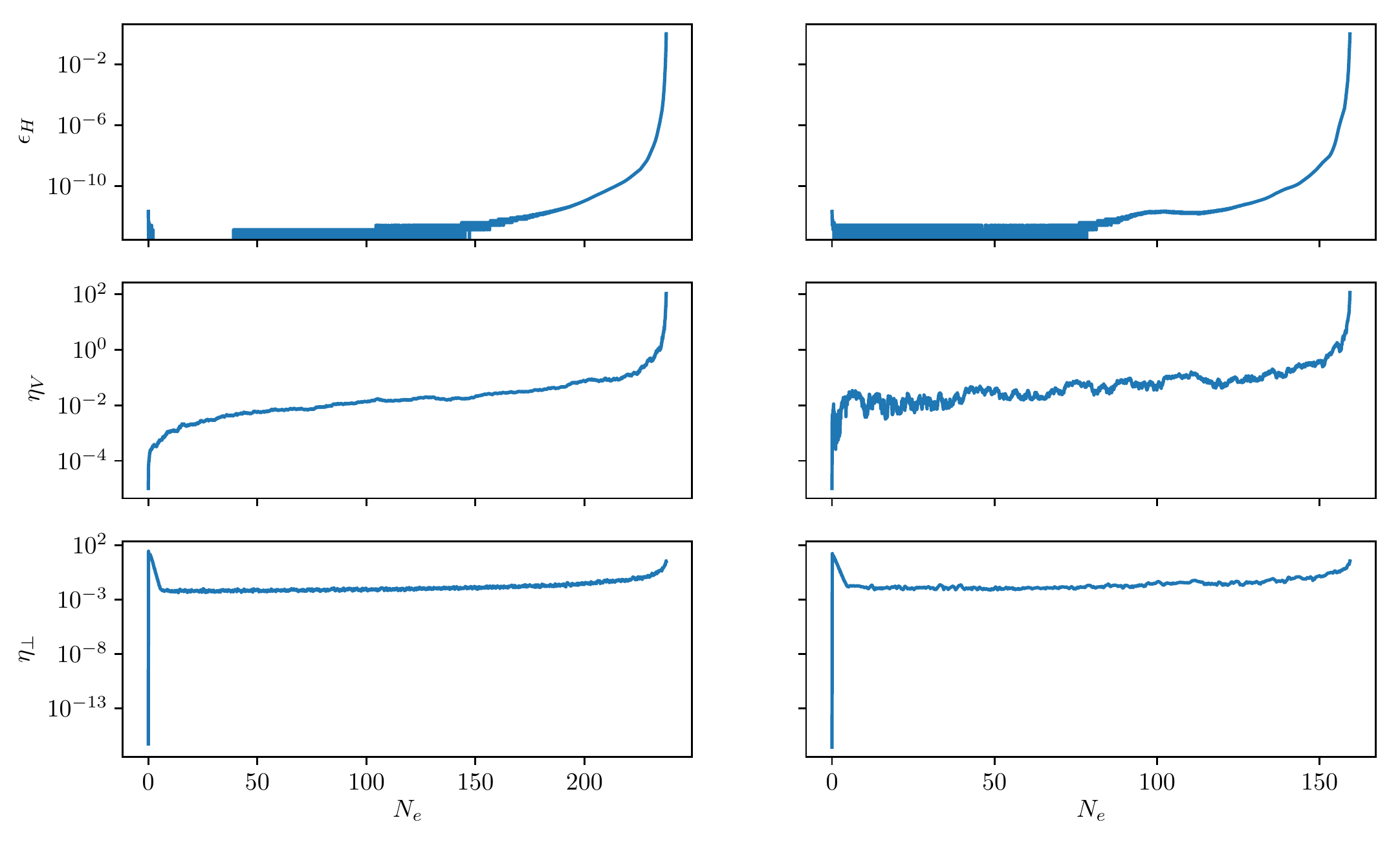}
    \caption{Some slow-roll parameters plotted against number of efoldings, for the typical realization (left), and atypical one (right).}
    \label{fig:single_sr}
\end{figure}

Some slow-roll parameters during the evolution are presented in Figure \ref{fig:single_sr}.
Inflation is approximately slow-roll (meaning $\eta$ and $\epsilon$ small) at early and intermediate times.
The late-time evolution is out of slow-roll, with $|\eta_V|>1$.
This latter period of inflation gives few efoldings, but covers the majority of the arclength.
$\eta_\perp$ (see (\ref{eq:etaperp})) is typically large twice during each realization, once when the initial field velocity isn't aligned with the lowest mass direction, and near the end of inflation as tachyonic directions open up.

\begin{figure}[h]
    \centering
    \includegraphics[width=0.9\textwidth]{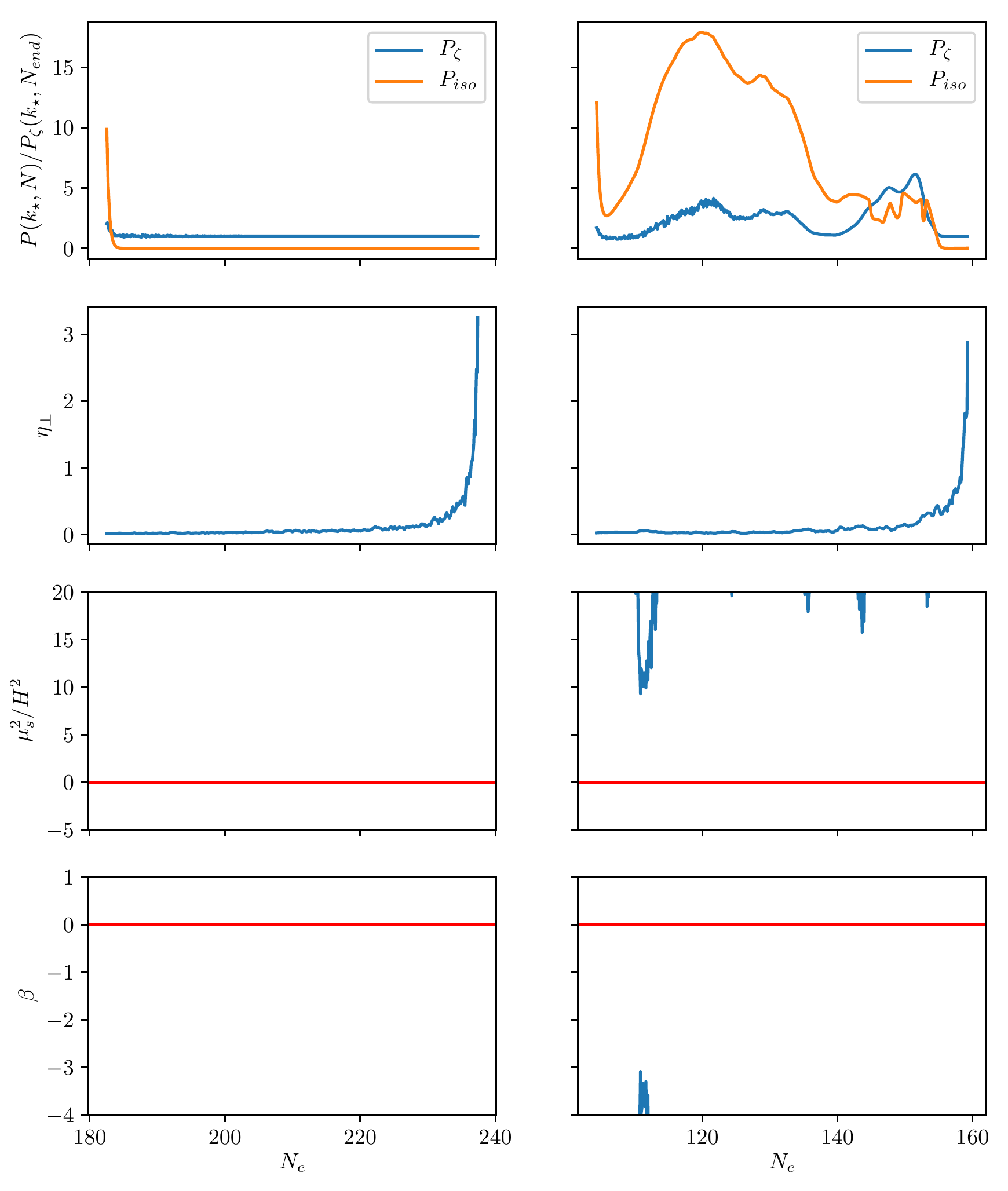}
    \caption{Isocurvature evolution and relevant parameters over the last 55 efolds of inflation, for the typical (left) and atypical (right) realizations. The estimate of first isocurvature mode self-coupling $\beta$ (cf. (\ref{eq:RSevol})), is never positive in the high-isocurvature realization. The first entropic mass $\mu_s^2$ is defined as $\mathcal{M}_{ss} + 3 \omega^2$ where $\omega^2$ measures the turning rate of the classical trajectory (cf. \ref{eq:defOmega}).}
    \label{fig:single_iso}
\end{figure}

\begin{figure}[H]
    \centering
    \includegraphics[width=0.6\textwidth]{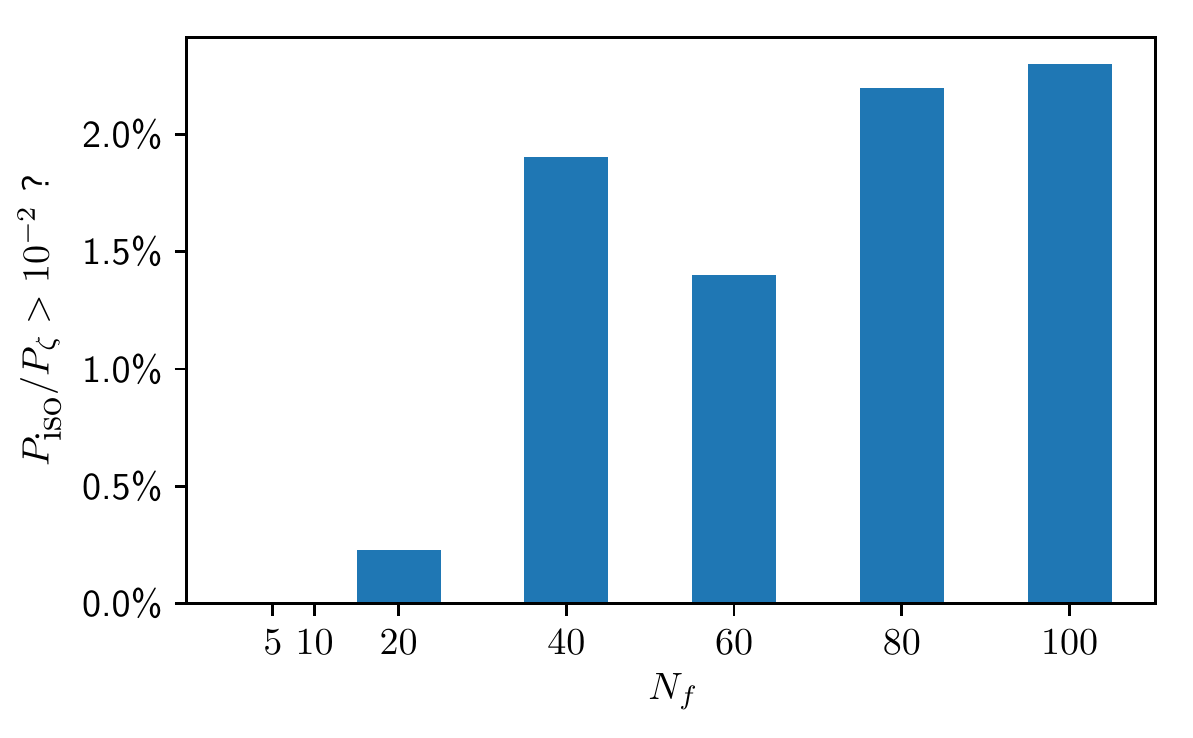}
    \caption{The percentage of realizations with an isocurvature power greater than 1\% of the adiabatic power at the end of inflation.}
    \label{fig:iso_percent}
\end{figure}

Power spectra and relevant parameters are presented in Figure \ref{fig:single_iso}.
The first isocurvature mass is positive for both realizations, and the first isocurvature mode self-coupling $\beta$ (see (\ref{eq:RSevol})), is negative for all times.
Despite the indicators of isocurvature self-coupling indicating decay, we see the growth (and then decay) of isocurvature.
We believe higher isocurvature modes to be responsible for the superhorizon evolution of the adiabatic and isocurvature power spectra in this realization.
A positive $\beta$ occurs extremely rarely in our models with high isocurvature and more than $10$ fields.
For comparison purposes, we plot such a realization, where $\beta$ drives the isocurvature evolution, in Figure \ref{fig:pos_beta}.
We believe this mechanism to be present in $\mathcal{O}(1\%)$ of similar 80-field realizations (see figure \ref{fig:iso_percent}).
\FloatBarrier
\subsection{Aggregate data}
\label{sec:data}
\FloatBarrier

\begin{figure}[h]
    \centering
    \includegraphics[width=0.9\textwidth]{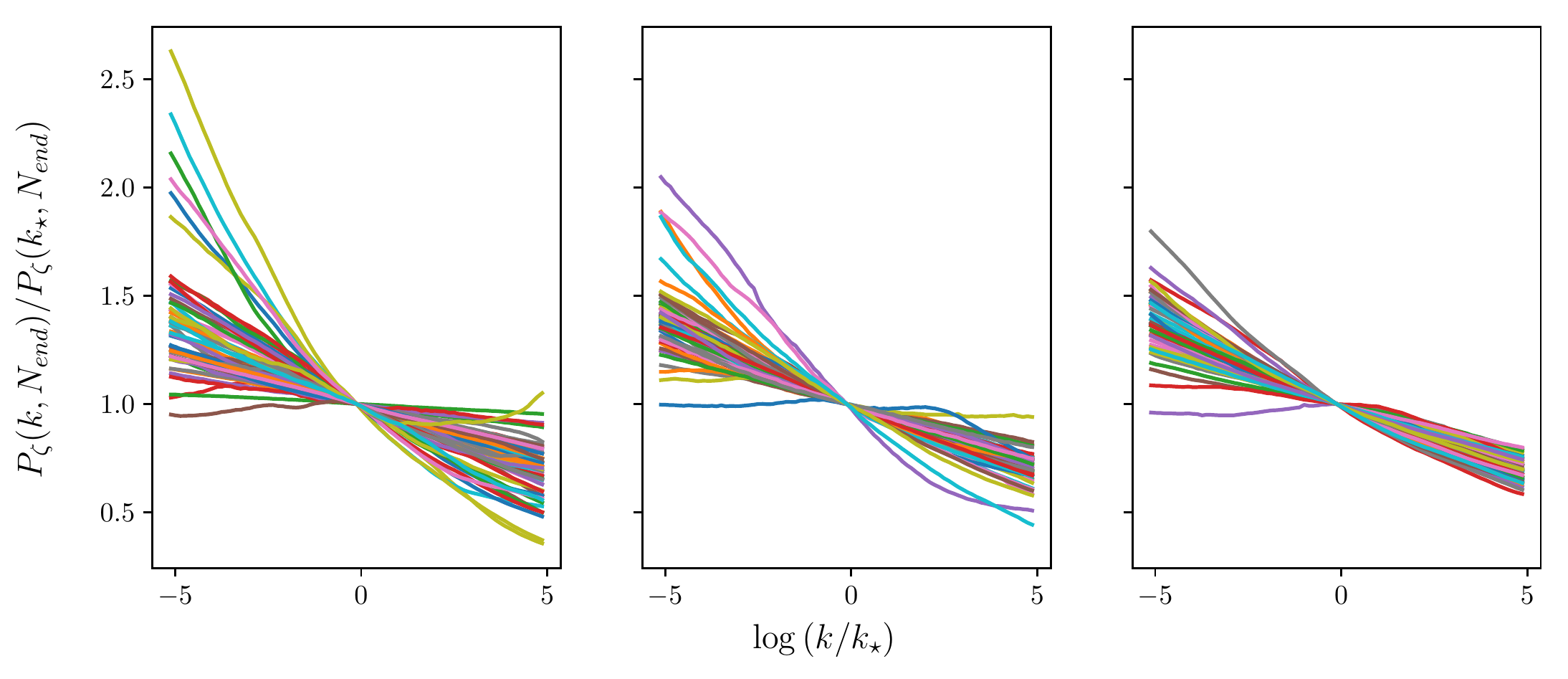}
    \caption{50 power spectra, for $k$ around the pivot scale. From left to right, these are 5, 40, and 80 fields. Drawn from dataset 1.}
    \label{fig:power spectra_Nf}
\end{figure}
In figure \ref{fig:power spectra_Nf}, we show a selection of power spectra as $N_f$ increases. Despite the potential being constructed patch-wise, the resulting power spectra are smooth and featureless.
At higher $N_f$, the power spectra become smoother and more predictive.This behavior was already observed and explained by Dias {\it et al.}  \cite{Dias:2017gva}. As $N_f$ increases, the range of masses increases only as $\sqrt{N_f}$ reducing the average separation between the masses. This effect  together with the characteristic eigenvalue repulsion of the DBM dynamics decreases the variability of the masses as $N_f$ grows. 
A few realizations with significant running of the spectral index are visible.

\begin{figure}[h!]
    \centering
    \includegraphics[width=0.95\textwidth]{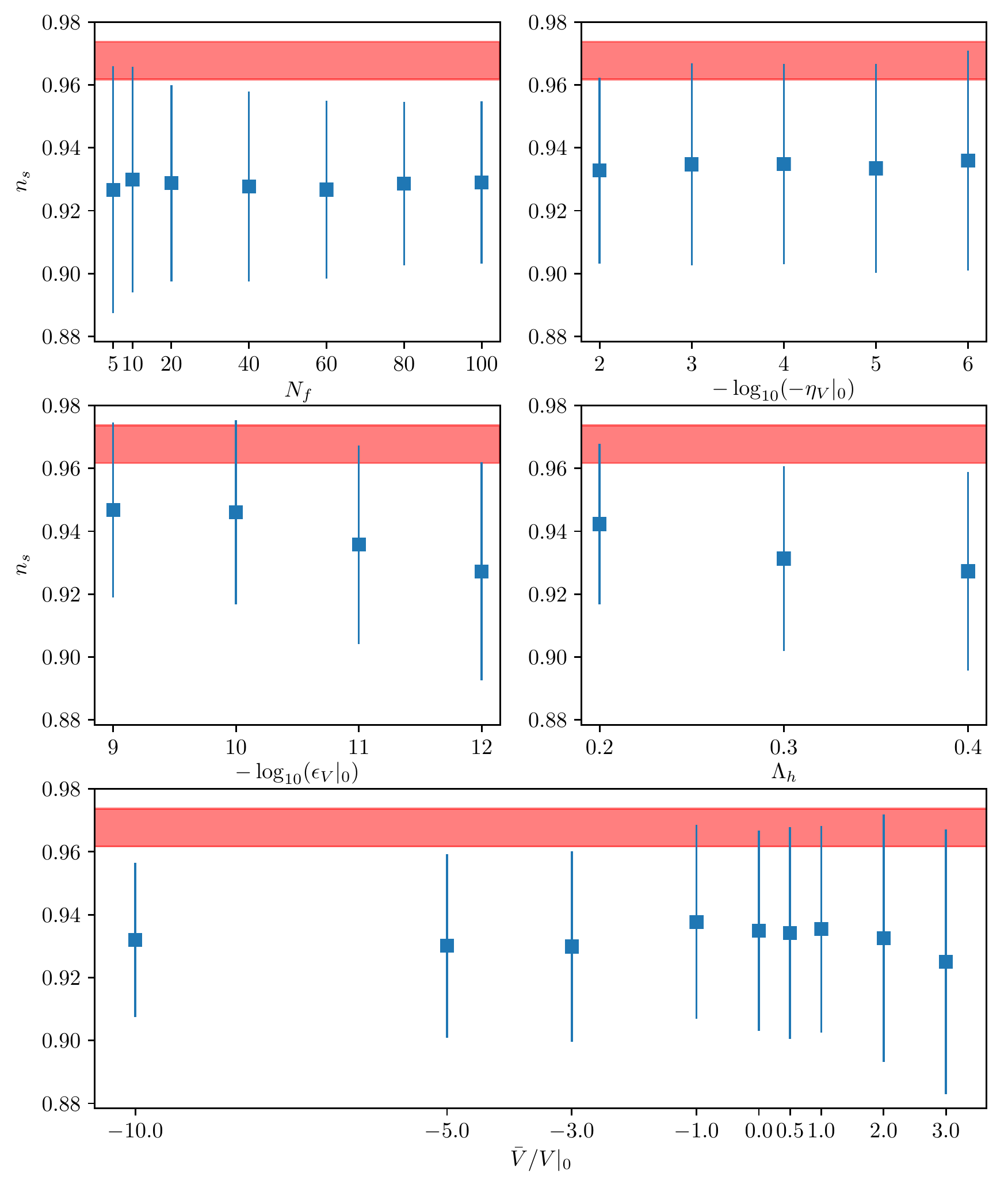}
    \caption{$n_s$ through various slices in parameter space. The red shaded region is the $1\sigma$ limits on $n_s$ from Planck \cite{Ade:2015lrj}.The plotted datasets are from Table \ref{tab:data}, with error bars showing $1\sigma$ width of the distribution. Each point on the figure corresponds to 1000 realizations.}
    \label{fig:ns_slices}
\end{figure}
\begin{figure}[H]
    \centering
    \includegraphics[width=0.8\textwidth]{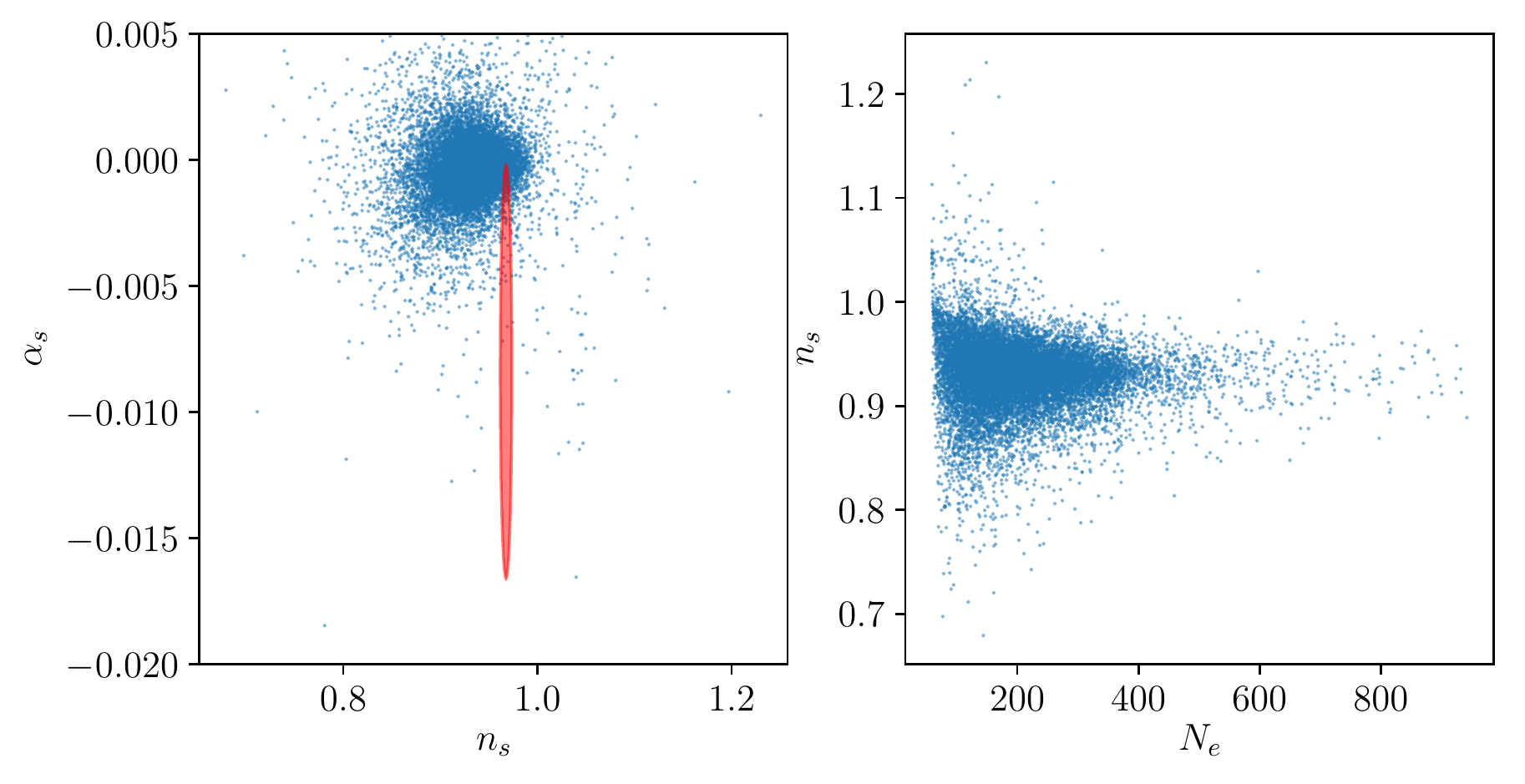}
    \caption{(left) The adiabatic observables for all datasets stacked, the red ellipse shows $1\sigma$ bounds from Planck. It's possible that improved bounds from Planck could rule out a more significant fraction of these models. (right) How $n_s$ depends on the number of efoldings.}
    \label{fig:ns_as_planck}
\end{figure}
\begin{figure}[H]
	\centering
	\includegraphics[width=0.45\textwidth]{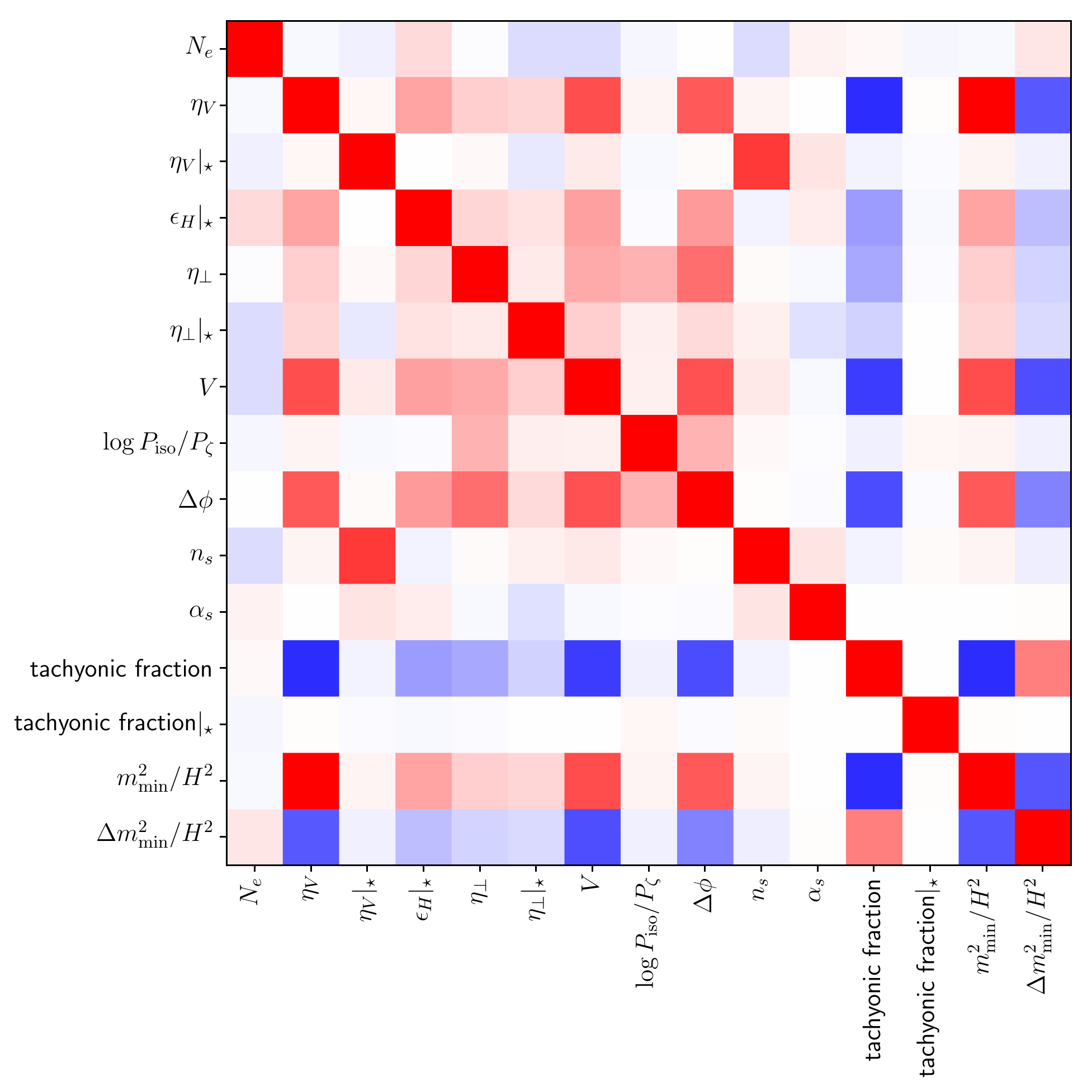}
	\includegraphics[width=0.45\textwidth]{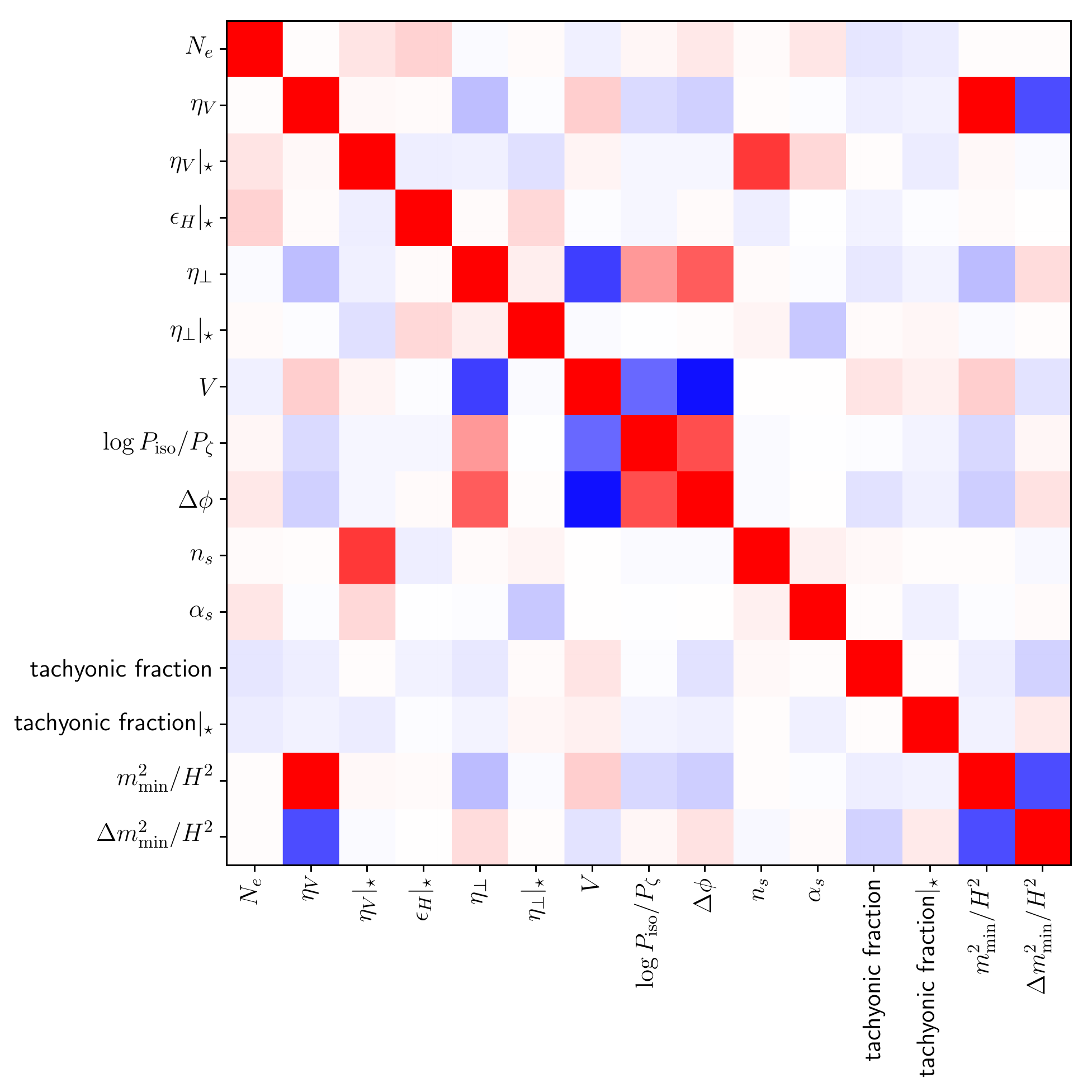}
	\includegraphics[width=0.65\textwidth]{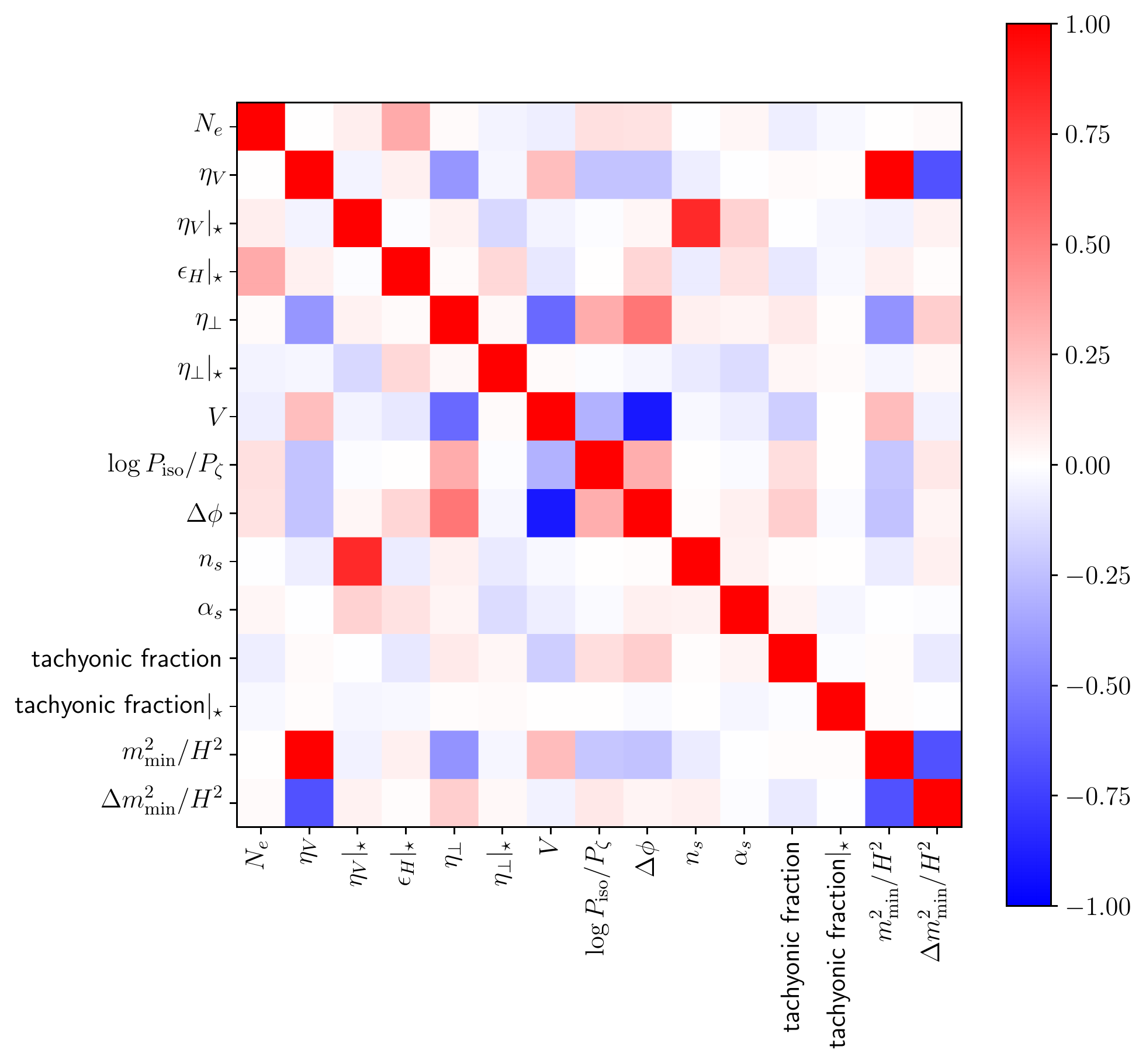}
	\caption{Linear correlation matrices for several relevant variables, for negative $V_0-\bar{V}$ (top left) positive (top right) and zero (center). Here, a pure red matrix square denotes a perfect positive correlation, and a pure blue square a perfect negative correlation. A decreased opacity denotes a less strong correlation, with white squares denoting uncorrelated variables. All variables with a $|_\star$ were measured when the pivot-scale mode exited the horizon. Variables marked with $|_0$ were measured at the start of inflation, and all other variables were measured at the end of inflation.}
	\label{fig:corr_matrix}
\end{figure}

\newpage
In figure \ref{fig:ns_slices}, we show how our parameter selections affect $n_s$.
The spectral index is strongly sensitive to the steepness of the initial inflationary point, $\epsilon_V|_0$, and the potential correlation length, $\Lambda_h$.
A steeper initial slope of the potential or a shorter correlation length both give a more blue $n_s$.
No other parameters, had a strong effect on $n_s$.
See our correlation plot, figure \ref{fig:corr_matrix}, for all parameters which affected $n_s$.
Most of our data was taken with an initial $\epsilon_V$ of $10^{-12}$, so most of our results have a too-red mean $n_s$ compared to Planck. We stress that these are mean results, and there is significant scatter around the mean. As we show in figure \ref{fig:ns_as_planck}, several thousand of our models are Planck-compatible.

\begin{figure}[H]
    \centering
    \includegraphics[width=0.95\textwidth]{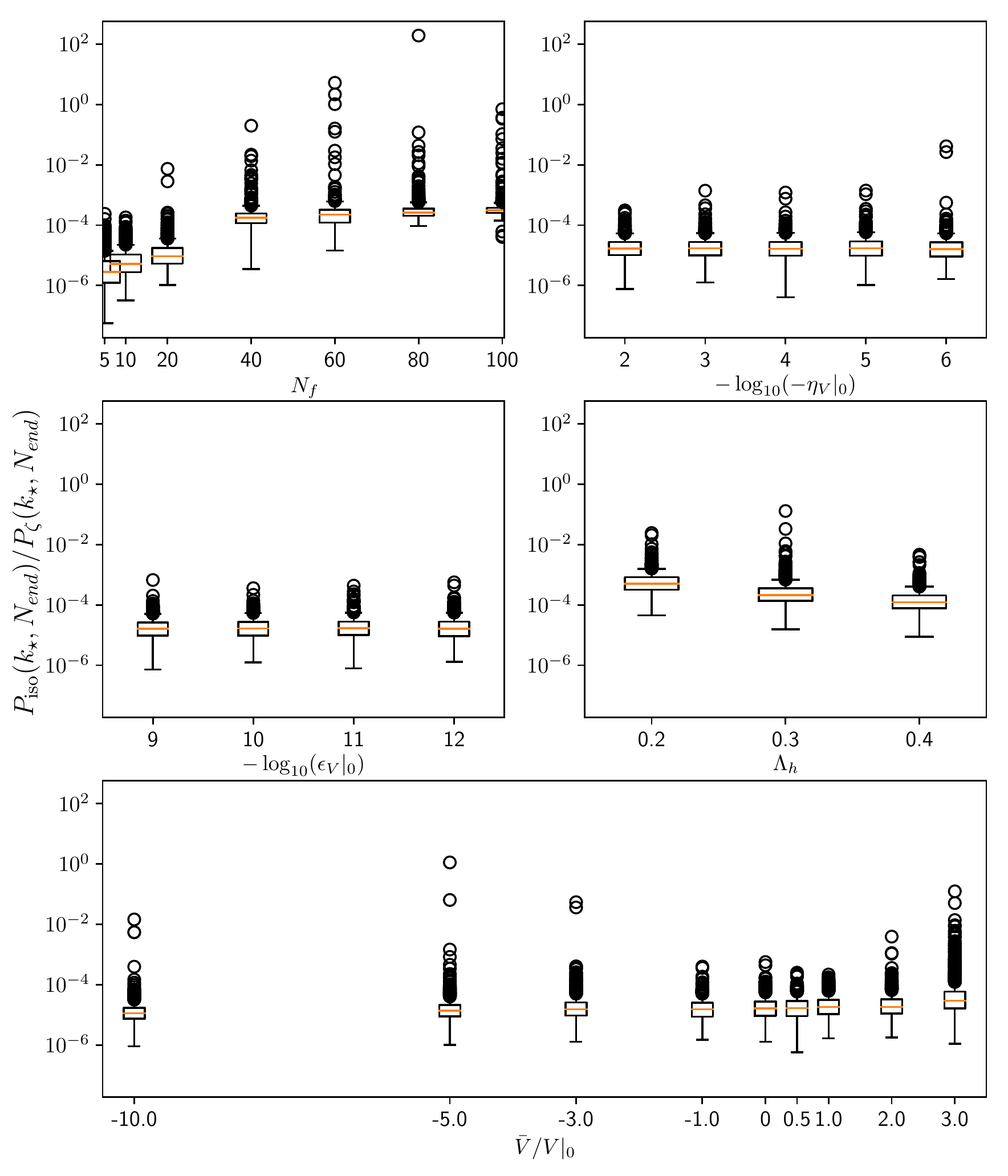}
    \caption{The pivot-scale isocurvature power through various slices in parameter space. Each box and whiskers corresponds to 1000 realizations. The boxes extend from the first quartile to the third, with the median marked as an orange line. Outliers (beyond $1.5$ times the difference of third and first quartiles) are marked with hollow circles. We note that all plotted parameter ranges have realizations with outlying isocurvature power, several significantly high $P_\text{iso}/P_\zeta \gtrsim 10^{-2}$.}
    \label{fig:piso_slices}
\end{figure}

In figure \ref{fig:piso_slices}, we show how our parameter selections affect the isocurvature to adiabatic power ratio at the end of inflation.
The strongest dependence on the median isocurvature comes from the number of fields, $N_f$.
In almost all datasets, there are significant outliers (defined in figure caption).
This is one of the significant results of this paper.
Not all realizations have ample time for their isocurvature to decay before the end of inflation. Following the argument in section \ref{sec:perturbations}, this is expected whenever one of the isocurvature modes' masses becomes tachyonic.

\begin{figure}[H]
    \centering
    \includegraphics[scale=0.8]{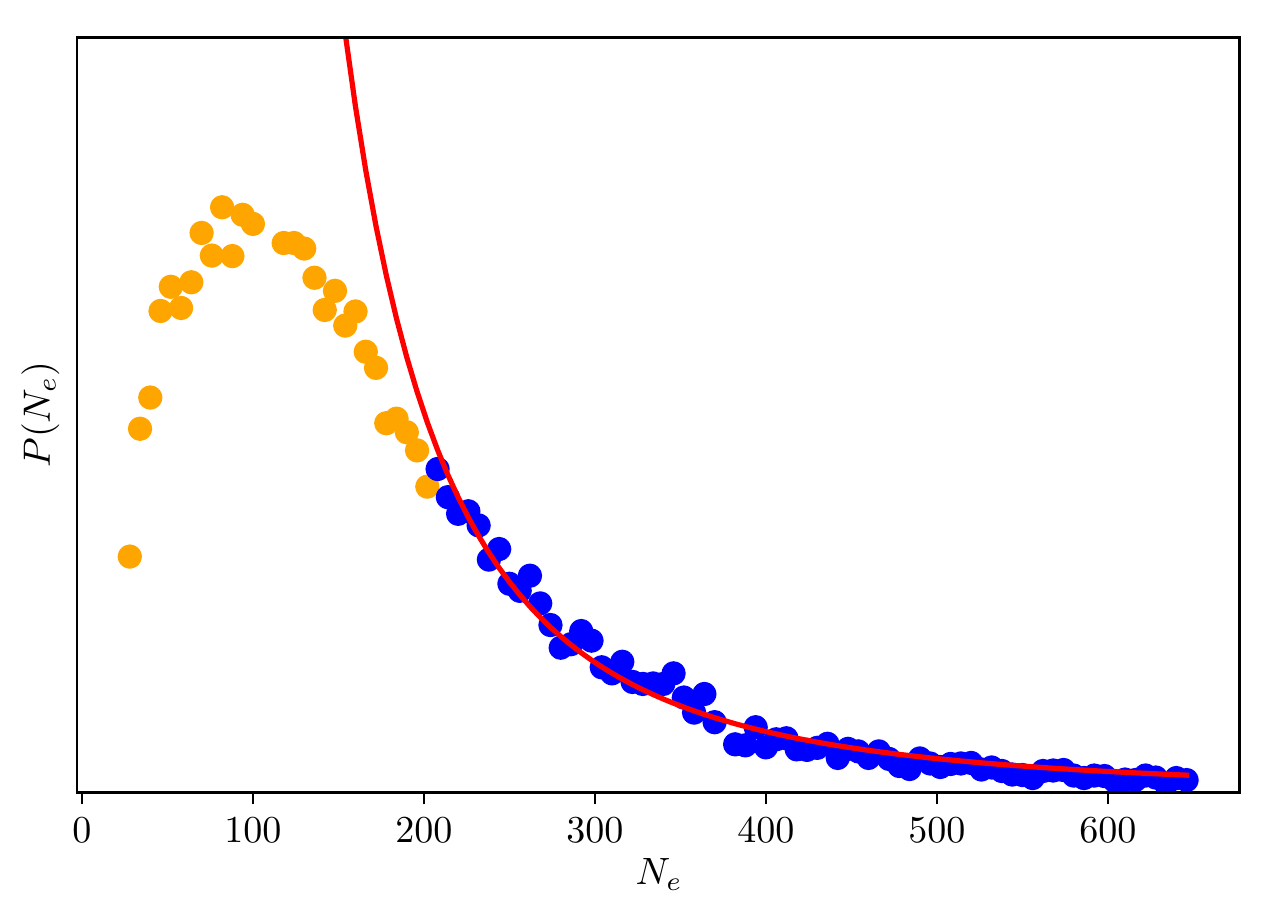}
    \caption{Masoumi et al. \cite{Masoumi:2016eag} predict a RGF inflating at saddle points has an approximate power-law e-fold distribution $P(N_e) \propto N_e^\gamma$, with $\gamma = -3$. Here, we plot our empirical e-fold probability from all datasets. The points in orange were excluded from the fit. The power law fit prefers $\gamma = -2.65 \pm 0.054$.}
    \label{fig:probNe}
\end{figure}

In figure \ref{fig:ns_as_planck}, we stack all of our adiabatic observables.
The distribution of the points should not be taken to be a fair sampling of the predictions of Gaussian DBM.
We chose most of our hyperparameters by hand, rather than having them determined by a distribution predicted by the model as in \cite{Masoumi:2017xbe,Masoumi:2016eag}.
Indeed, it's unclear if locally-defined potentials can fairly sample the landscape without comparison to some globally defined construction -- the initial conditions must be chosen by hand.
Most of our realizations were taken in an unfavorable regime of parameter space with $\epsilon_V|_0 = 10^{-12}$, so most of these data points have a too-red $n_s$.
Nevertheless, several hundred realizations fall within the Planck $1\sigma$ ellipse. 
It's possible that tighter constraints from Planck will eliminate this class of models.
Due to the stochastic nature of our model, it's likely that some vanishingly small fraction of realizations fit any reasonable tightening of the Planck results.
In the right half of the figure, we show how $n_s$ depends on the number of efolds.
Most realizations with a Planck-compatible $n_s$ have $\approx 200$ efolds.

In figure \ref{fig:corr_matrix}, we present a cross-correlation matrix of several background and perturbative quantities.
We use only $N_f=10$ data, and split it by sign of $V_0-\bar{V}$.
In all correlation matrices, we see that $n_s$ correlates with $\eta_V$ at horizon exit, but not $\epsilon_H$. This is expected, since $\left| \eta_V \right| \gg \epsilon_V$ near the pivot scale.
The isocurvature power seems to correlate strongly with $\eta_\perp$ for all signs of $V_0-\bar{V}$. From our argument in section \ref{sec:perturbations}, a large $\omega$ will allow the first isocurvature mode to decay, while giving the higher isocurvature modes a positive contribution to their masses squared. Some nonzero $\omega$ is however necessary to source isocurvature.
$\eta_V$ correlates with $V$ when $V-\bar{V} < 0$, but anticorrelates when $V-\bar{V}$ is positive or zero.
This is consistent with the behavior of the equilibrium mass distribution (\ref{eq:shiftedWigner}).

\FloatBarrier
\section{Conclusions}

We presented an analysis of multi-field inflation in random potentials generated via a modified Dyson-Brownian dynamics that includes correlations between the Hessian and the height of the potential like a Random Gaussian Field.
These models have a rich parameter space, Planck compatibility (reheating caveats notwithstanding) is not rare though  $\sim 1-2 \% $ of the cases have unobserved high isocurvature power spectra. The correlations between the observables $n_s$, $\alpha_s$ and ${\cal P}_{\rm{iso}}/{\cal P}_{\zeta}$ and background quantities are best summarized in the cross-correlation matrix presented in fig \ref{fig:corr_matrix}. How changing hyperparameters affects the observables is best summarized by figures \ref{fig:ns_slices} and \ref{fig:piso_slices}.

Broadly these results agree with previous work in multi-field inflation in RGFs \cite{Masoumi:2017xbe,Bjorkmo:2018txh} in finding that, regardless of the value of $N_f$, Planck compatibility is not rare when the initial conditions are chosen to guarantee, at least, 55 e-foldings of inflation. The results are also consistent with Dias {\it et al.} \cite{Dias:2017gva}, in spite of the modified evolution. In our models, as the number of fields increases so does the power in isocurvature modes, but the percentage of cases when the Planck bound is exceeded is below $2 \%$ for the largest $N_f=100$. This small percentage of cases seems consistent with Bjorkmo and Marsh \cite{Bjorkmo:2018txh} though we are unable to compare the exact percentage. There is no inconsistency with Masoumi{\it et al.} \cite{Masoumi:2017xbe} either since they predict a single field behavior whenever $\Lambda_h N_f^{-1/4} \ll 1$ and the values of $\Lambda_h$ and $N_f$ that we considered in this work do not satisfy this bound.

\section{Acknowledgements}

It is a pleasure to thank T. Bachlechner, M. Dias, J. Frazer, D. Marsh, A. Masoumi, A. Vilenkin  and M. Yamada for interesting discussions. In addition, we thank D. Marsh, A. Valenkin and M. Yamada for feedback on the manuscript. Part of this work was done at Aspen Center for Physics, which is supported by U.S. National Science Foundation
grant PHY-1607611. This work was also supported by the U.S. National Science Foundation under Grants PHY–1521186 and PHY–1620610. The authors acknowledge the Texas Advanced Computing Center (TACC) at The University of Texas at Austin for providing HPC resources that have contributed to the research results reported within this paper. 
	
\bibliography{refs2}{}

\providecommand{\href}[2]{#2}\begingroup\raggedright\begin{thebibliography}{10}

\bibitem{Ade:2015lrj}
{\scshape Planck} collaboration, P.~A.~R. Ade et~al., \emph{{Planck 2015
  results. XX. Constraints on inflation}},
  \href{https://doi.org/10.1051/0004-6361/201525898}{\emph{Astron. Astrophys.}
  {\bfseries 594} (2016) A20}
  [\href{https://arxiv.org/abs/1502.02114}{{\ttfamily 1502.02114}}].

\bibitem{Mollerach:1989hu}
S.~Mollerach, \emph{{Isocurvature Baryon Perturbations and Inflation}},
  \href{https://doi.org/10.1103/PhysRevD.42.313}{\emph{Phys. Rev.} {\bfseries
  D42} (1990) 313}.

\bibitem{Weinberg:2004kf}
S.~Weinberg, \emph{{Must cosmological perturbations remain non-adiabatic after
  multi-field inflation?}},
  \href{https://doi.org/10.1103/PhysRevD.70.083522}{\emph{Phys. Rev.}
  {\bfseries D70} (2004) 083522}
  [\href{https://arxiv.org/abs/astro-ph/0405397}{{\ttfamily
  astro-ph/0405397}}].

\bibitem{Hotinli:2017vhx}
S.~C. Hotinli, J.~Frazer, A.~H. Jaffe, J.~Meyers, L.~C. Price and E.~R.~M.
  Tarrant, \emph{{Effect of reheating on predictions following multiple-field
  inflation}}, \href{https://doi.org/10.1103/PhysRevD.97.023511}{\emph{Phys.
  Rev.} {\bfseries D97} (2018) 023511}
  [\href{https://arxiv.org/abs/1710.08913}{{\ttfamily 1710.08913}}].

\bibitem{Lyth:1984gv}
D.~H. Lyth, \emph{{Large Scale Energy Density Perturbations and Inflation}},
  \href{https://doi.org/10.1103/PhysRevD.31.1792}{\emph{Phys. Rev.} {\bfseries
  D31} (1985) 1792}.

\bibitem{Starobinsky:1986fxa}
A.~A. Starobinsky, \emph{{Multicomponent de Sitter (Inflationary) Stages and
  the Generation of Perturbations}}, {\emph{JETP Lett.} {\bfseries 42} (1985)
  152}.

\bibitem{Salopek:1990jq}
D.~S. Salopek and J.~R. Bond, \emph{{Nonlinear evolution of long wavelength
  metric fluctuations in inflationary models}},
  \href{https://doi.org/10.1103/PhysRevD.42.3936}{\emph{Phys. Rev.} {\bfseries
  D42} (1990) 3936}.

\bibitem{Sasaki:1995aw}
M.~Sasaki and E.~D. Stewart, \emph{{A General analytic formula for the spectral
  index of the density perturbations produced during inflation}},
  \href{https://doi.org/10.1143/PTP.95.71}{\emph{Prog. Theor. Phys.} {\bfseries
  95} (1996) 71} [\href{https://arxiv.org/abs/astro-ph/9507001}{{\ttfamily
  astro-ph/9507001}}].

\bibitem{Sasaki:1998ug}
M.~Sasaki and T.~Tanaka, \emph{{Superhorizon scale dynamics of multiscalar
  inflation}}, \href{https://doi.org/10.1143/PTP.99.763}{\emph{Prog. Theor.
  Phys.} {\bfseries 99} (1998) 763}
  [\href{https://arxiv.org/abs/gr-qc/9801017}{{\ttfamily gr-qc/9801017}}].

\bibitem{GrootNibbelink:2000vx}
S.~Groot~Nibbelink and B.~J.~W. van Tent, \emph{{Density perturbations arising
  from multiple field slow roll inflation}},
  \href{https://arxiv.org/abs/hep-ph/0011325}{{\ttfamily hep-ph/0011325}}.

\bibitem{GrootNibbelink:2001qt}
S.~Groot~Nibbelink and B.~J.~W. van Tent, \emph{{Scalar perturbations during
  multiple field slow-roll inflation}},
  \href{https://doi.org/10.1088/0264-9381/19/4/302}{\emph{Class. Quant. Grav.}
  {\bfseries 19} (2002) 613}
  [\href{https://arxiv.org/abs/hep-ph/0107272}{{\ttfamily hep-ph/0107272}}].

\bibitem{Wands:2000dp}
D.~Wands, K.~A. Malik, D.~H. Lyth and A.~R. Liddle, \emph{{A New approach to
  the evolution of cosmological perturbations on large scales}},
  \href{https://doi.org/10.1103/PhysRevD.62.043527}{\emph{Phys. Rev.}
  {\bfseries D62} (2000) 043527}
  [\href{https://arxiv.org/abs/astro-ph/0003278}{{\ttfamily
  astro-ph/0003278}}].

\bibitem{Amendola:2001ni}
L.~Amendola, C.~Gordon, D.~Wands and M.~Sasaki, \emph{{Correlated perturbations
  from inflation and the cosmic microwave background}},
  \href{https://doi.org/10.1103/PhysRevLett.88.211302}{\emph{Phys. Rev. Lett.}
  {\bfseries 88} (2002) 211302}
  [\href{https://arxiv.org/abs/astro-ph/0107089}{{\ttfamily
  astro-ph/0107089}}].

\bibitem{Rigopoulos:2004gr}
G.~I. Rigopoulos and E.~P.~S. Shellard, \emph{{Non-linear inflationary
  perturbations}},
  \href{https://doi.org/10.1088/1475-7516/2005/10/006}{\emph{JCAP} {\bfseries
  0510} (2005) 006} [\href{https://arxiv.org/abs/astro-ph/0405185}{{\ttfamily
  astro-ph/0405185}}].

\bibitem{Rigopoulos:2005xx}
G.~I. Rigopoulos, E.~P.~S. Shellard and B.~J.~W. van Tent, \emph{{Non-linear
  perturbations in multiple-field inflation}},
  \href{https://doi.org/10.1103/PhysRevD.73.083521}{\emph{Phys. Rev.}
  {\bfseries D73} (2006) 083521}
  [\href{https://arxiv.org/abs/astro-ph/0504508}{{\ttfamily
  astro-ph/0504508}}].

\bibitem{Yokoyama:2007uu}
S.~Yokoyama, T.~Suyama and T.~Tanaka, \emph{{Primordial Non-Gaussianity in
  Multi-Scalar Slow-Roll Inflation}},
  \href{https://doi.org/10.1088/1475-7516/2007/07/013}{\emph{JCAP} {\bfseries
  0707} (2007) 013} [\href{https://arxiv.org/abs/0705.3178}{{\ttfamily
  0705.3178}}].

\bibitem{Lalak:2007vi}
Z.~Lalak, D.~Langlois, S.~Pokorski and K.~Turzynski, \emph{{Curvature and
  isocurvature perturbations in two-field inflation}},
  \href{https://doi.org/10.1088/1475-7516/2007/07/014}{\emph{JCAP} {\bfseries
  0707} (2007) 014} [\href{https://arxiv.org/abs/0704.0212}{{\ttfamily
  0704.0212}}].

\bibitem{Ringeval:2007am}
C.~Ringeval, \emph{{The exact numerical treatment of inflationary models}},
  \href{https://doi.org/10.1007/978-3-540-74353-8_7}{\emph{Lect. Notes Phys.}
  {\bfseries 738} (2008) 243}
  [\href{https://arxiv.org/abs/astro-ph/0703486}{{\ttfamily
  astro-ph/0703486}}].

\bibitem{Yokoyama:2008by}
S.~Yokoyama, T.~Suyama and T.~Tanaka, \emph{{Efficient diagrammatic computation
  method for higher order correlation functions of local type primordial
  curvature perturbations}},
  \href{https://doi.org/10.1088/1475-7516/2009/02/012}{\emph{JCAP} {\bfseries
  0902} (2009) 012} [\href{https://arxiv.org/abs/0810.3053}{{\ttfamily
  0810.3053}}].

\bibitem{Malik:2008im}
K.~A. Malik and D.~Wands, \emph{{Cosmological perturbations}},
  \href{https://doi.org/10.1016/j.physrep.2009.03.001}{\emph{Phys. Rept.}
  {\bfseries 475} (2009) 1} [\href{https://arxiv.org/abs/0809.4944}{{\ttfamily
  0809.4944}}].

\bibitem{Mulryne:2009kh}
D.~J. Mulryne, D.~Seery and D.~Wesley, \emph{{Moment transport equations for
  non-Gaussianity}},
  \href{https://doi.org/10.1088/1475-7516/2010/01/024}{\emph{JCAP} {\bfseries
  1001} (2010) 024} [\href{https://arxiv.org/abs/0909.2256}{{\ttfamily
  0909.2256}}].

\bibitem{Chen:2009zp}
X.~Chen and Y.~Wang, \emph{{Quasi-Single Field Inflation and
  Non-Gaussianities}},
  \href{https://doi.org/10.1088/1475-7516/2010/04/027}{\emph{JCAP} {\bfseries
  1004} (2010) 027} [\href{https://arxiv.org/abs/0911.3380}{{\ttfamily
  0911.3380}}].

\bibitem{Mulryne:2010rp}
D.~J. Mulryne, D.~Seery and D.~Wesley, \emph{{Moment transport equations for
  the primordial curvature perturbation}},
  \href{https://doi.org/10.1088/1475-7516/2011/04/030}{\emph{JCAP} {\bfseries
  1104} (2011) 030} [\href{https://arxiv.org/abs/1008.3159}{{\ttfamily
  1008.3159}}].

\bibitem{Peterson:2010np}
C.~M. Peterson and M.~Tegmark, \emph{{Testing Two-Field Inflation}},
  \href{https://doi.org/10.1103/PhysRevD.83.023522}{\emph{Phys. Rev.}
  {\bfseries D83} (2011) 023522}
  [\href{https://arxiv.org/abs/1005.4056}{{\ttfamily 1005.4056}}].

\bibitem{Peterson:2011yt}
C.~M. Peterson and M.~Tegmark, \emph{{Testing multifield inflation: A geometric
  approach}}, \href{https://doi.org/10.1103/PhysRevD.87.103507}{\emph{Phys.
  Rev.} {\bfseries D87} (2013) 103507}
  [\href{https://arxiv.org/abs/1111.0927}{{\ttfamily 1111.0927}}].

\bibitem{Achucarro:2010da}
A.~Achucarro, J.-O. Gong, S.~Hardeman, G.~A. Palma and S.~P. Patil,
  \emph{{Features of heavy physics in the CMB power spectrum}},
  \href{https://doi.org/10.1088/1475-7516/2011/01/030}{\emph{JCAP} {\bfseries
  1101} (2011) 030} [\href{https://arxiv.org/abs/1010.3693}{{\ttfamily
  1010.3693}}].

\bibitem{Cremonini:2010ua}
S.~Cremonini, Z.~Lalak and K.~Turzynski, \emph{{Strongly Coupled Perturbations
  in Two-Field Inflationary Models}},
  \href{https://doi.org/10.1088/1475-7516/2011/03/016}{\emph{JCAP} {\bfseries
  1103} (2011) 016} [\href{https://arxiv.org/abs/1010.3021}{{\ttfamily
  1010.3021}}].

\bibitem{Avgoustidis:2011em}
A.~Avgoustidis, S.~Cremonini, A.-C. Davis, R.~H. Ribeiro, K.~Turzynski and
  S.~Watson, \emph{{The Importance of Slow-roll Corrections During Multi-field
  Inflation}}, \href{https://doi.org/10.1088/1475-7516/2012/02/038}{\emph{JCAP}
  {\bfseries 1202} (2012) 038}
  [\href{https://arxiv.org/abs/1110.4081}{{\ttfamily 1110.4081}}].

\bibitem{Lehners:2009ja}
J.-L. Lehners and S.~Renaux-Petel, \emph{{Multifield Cosmological Perturbations
  at Third Order and the Ekpyrotic Trispectrum}},
  \href{https://doi.org/10.1103/PhysRevD.80.063503}{\emph{Phys. Rev.}
  {\bfseries D80} (2009) 063503}
  [\href{https://arxiv.org/abs/0906.0530}{{\ttfamily 0906.0530}}].

\bibitem{Huston:2009ac}
I.~Huston and K.~A. Malik, \emph{{Numerical calculation of second order
  perturbations}},
  \href{https://doi.org/10.1088/1475-7516/2009/09/019}{\emph{JCAP} {\bfseries
  0909} (2009) 019} [\href{https://arxiv.org/abs/0907.2917}{{\ttfamily
  0907.2917}}].

\bibitem{Kaiser:2013sna}
D.~I. Kaiser and E.~I. Sfakianakis, \emph{{Multifield Inflation after Planck:
  The Case for Nonminimal Couplings}},
  \href{https://doi.org/10.1103/PhysRevLett.112.011302}{\emph{Phys. Rev. Lett.}
  {\bfseries 112} (2014) 011302}
  [\href{https://arxiv.org/abs/1304.0363}{{\ttfamily 1304.0363}}].

\bibitem{Schutz:2013fua}
K.~Schutz, E.~I. Sfakianakis and D.~I. Kaiser, \emph{{Multifield Inflation
  after Planck: Isocurvature Modes from Nonminimal Couplings}},
  \href{https://doi.org/10.1103/PhysRevD.89.064044}{\emph{Phys. Rev.}
  {\bfseries D89} (2014) 064044}
  [\href{https://arxiv.org/abs/1310.8285}{{\ttfamily 1310.8285}}].

\bibitem{Dias:2016slx}
M.~Dias, J.~Frazer and M.~C.~D. Marsh, \emph{{Simple emergent power spectra
  from complex inflationary physics}},
  \href{https://doi.org/10.1103/PhysRevLett.117.141303}{\emph{Phys. Rev. Lett.}
  {\bfseries 117} (2016) 141303}
  [\href{https://arxiv.org/abs/1604.05970}{{\ttfamily 1604.05970}}].

\bibitem{Dias:2017gva}
M.~Dias, J.~Frazer and M.~c.~D. Marsh, \emph{{Seven Lessons from Manyfield
  Inflation in Random Potentials}},
  \href{https://doi.org/10.1088/1475-7516/2018/01/036}{\emph{JCAP} {\bfseries
  1801} (2018) 036} [\href{https://arxiv.org/abs/1706.03774}{{\ttfamily
  1706.03774}}].

\bibitem{Masoumi:2016eag}
A.~Masoumi, A.~Vilenkin and M.~Yamada, \emph{{Inflation in random Gaussian
  landscapes}},
  \href{https://doi.org/10.1088/1475-7516/2017/05/053}{\emph{JCAP} {\bfseries
  1705} (2017) 053} [\href{https://arxiv.org/abs/1612.03960}{{\ttfamily
  1612.03960}}].

\bibitem{Masoumi:2017gmh}
A.~Masoumi, A.~Vilenkin and M.~Yamada, \emph{{Initial conditions for slow-roll
  inflation in a random Gaussian landscape}},
  \href{https://doi.org/10.1088/1475-7516/2017/07/003}{\emph{JCAP} {\bfseries
  1707} (2017) 003} [\href{https://arxiv.org/abs/1704.06994}{{\ttfamily
  1704.06994}}].

\bibitem{Masoumi:2017xbe}
A.~Masoumi, A.~Vilenkin and M.~Yamada, \emph{{Inflation in multi-field random
  Gaussian landscapes}},
  \href{https://doi.org/10.1088/1475-7516/2017/12/035}{\emph{JCAP} {\bfseries
  1712} (2017) 035} [\href{https://arxiv.org/abs/1707.03520}{{\ttfamily
  1707.03520}}].

\bibitem{Freivogel:2016kxc}
B.~Freivogel, R.~Gobbetti, E.~Pajer and I.-S. Yang, \emph{{Inflation on a
  Slippery Slope}},  \href{https://arxiv.org/abs/1608.00041}{{\ttfamily
  1608.00041}}.

\bibitem{Bjorkmo:2017nzd}
T.~Bjorkmo and M.~C.~D. Marsh, \emph{{Manyfield Inflation in Random
  Potentials}},  \href{https://arxiv.org/abs/1709.10076}{{\ttfamily
  1709.10076}}.

\bibitem{Bjorkmo:2018txh}
T.~Bjorkmo and M.~C.~D. Marsh, \emph{{Local, algebraic simplifications of
  Gaussian random fields}},  \href{https://arxiv.org/abs/1805.03117}{{\ttfamily
  1805.03117}}.

\bibitem{Achucarro:2017ing}
A.~Achúcarro, R.~Kallosh, A.~Linde, D.-G. Wang and Y.~Welling,
  \emph{{Universality of multi-field $\alpha$-attractors}},
  \href{https://doi.org/10.1088/1475-7516/2018/04/028}{\emph{JCAP} {\bfseries
  1804} (2018) 028} [\href{https://arxiv.org/abs/1711.09478}{{\ttfamily
  1711.09478}}].

\bibitem{Leach:2001zf}
S.~M. Leach, M.~Sasaki, D.~Wands and A.~R. Liddle, \emph{{Enhancement of
  superhorizon scale inflationary curvature perturbations}},
  \href{https://doi.org/10.1103/PhysRevD.64.023512}{\emph{Phys. Rev.}
  {\bfseries D64} (2001) 023512}
  [\href{https://arxiv.org/abs/astro-ph/0101406}{{\ttfamily
  astro-ph/0101406}}].

\bibitem{Marsh:2013qca}
M.~C.~D. Marsh, L.~McAllister, E.~Pajer and T.~Wrase, \emph{{Charting an
  Inflationary Landscape with Random Matrix Theory}},
  \href{https://doi.org/10.1088/1475-7516/2013/11/040}{\emph{JCAP} {\bfseries
  1311} (2013) 040} [\href{https://arxiv.org/abs/1307.3559}{{\ttfamily
  1307.3559}}].

\bibitem{Yamada:2017uzq}
M.~Yamada and A.~Vilenkin, \emph{{Hessian eigenvalue distribution in a random
  Gaussian landscape}},  \href{https://arxiv.org/abs/1712.01282}{{\ttfamily
  1712.01282}}.

\bibitem{Bachlechner:2014rqa}
T.~C. Bachlechner, \emph{{On Gaussian Random Supergravity}},
  \href{https://doi.org/10.1007/JHEP04(2014)054}{\emph{JHEP} {\bfseries 04}
  (2014) 054} [\href{https://arxiv.org/abs/1401.6187}{{\ttfamily 1401.6187}}].

\bibitem{Dias:2015rca}
M.~Dias, J.~Frazer and D.~Seery, \emph{{Computing observables in curved
  multifield models of inflation—A guide (with code) to the transport
  method}}, \href{https://doi.org/10.1088/1475-7516/2015/12/030}{\emph{JCAP}
  {\bfseries 1512} (2015) 030}
  [\href{https://arxiv.org/abs/1502.03125}{{\ttfamily 1502.03125}}].

\bibitem{Mehta1991}
M.~L. Mehta, \emph{Random Matrices}. Academic Press, New York, 1st~ed., 1991.

\bibitem{Mehta2004}
M.~L. Mehta, \emph{Random Matrices}. Academic Press, New York, 3rd~ed., 2004.

\bibitem{doi:10.1063/1.1703862}
F.~J. Dyson, \emph{A brownian‐motion model for the eigenvalues of a random
  matrix}, \href{https://doi.org/10.1063/1.1703862}{\emph{Journal of
  Mathematical Physics} {\bfseries 3} (1962) 1191}
  [\href{https://arxiv.org/abs/https://doi.org/10.1063/1.1703862}{{\ttfamily
  https://doi.org/10.1063/1.1703862}}].

\bibitem{Easther:2016ire}
R.~Easther, A.~H. Guth and A.~Masoumi, \emph{{Counting Vacua in Random
  Landscapes}},  \href{https://arxiv.org/abs/1612.05224}{{\ttfamily
  1612.05224}}.

\bibitem{Wang:2016kzp}
G.~Wang and T.~Battefeld, \emph{{Random Functions via Dyson Brownian Motion:
  Progress and Problems}},
  \href{https://doi.org/10.1088/1475-7516/2016/09/008}{\emph{JCAP} {\bfseries
  1609} (2016) 008} [\href{https://arxiv.org/abs/1607.02514}{{\ttfamily
  1607.02514}}].

\bibitem{Fyodorov:2004}
Y.~V. Fyodorov, \emph{Complexity of random energy landscapes, glass transition,
  and absolute value of the spectral determinant of random matrices},
  \href{https://doi.org/10.1103/PhysRevLett.92.240601}{\emph{Phys. Rev. Lett.}
  {\bfseries 92} (2004) 240601}.

\bibitem{Bray:2007tf}
A.~J. Bray and D.~S. Dean, \emph{{Statistics of critical points of Gaussian
  fields on large-dimensional spaces}},
  \href{https://doi.org/10.1103/PhysRevLett.98.150201}{\emph{Phys. Rev. Lett.}
  {\bfseries 98} (2007) 150201}.

\bibitem{Battefeld:2014qoa}
T.~Battefeld and C.~Modi, \emph{{Local random potentials of high
  differentiability to model the Landscape}},
  \href{https://doi.org/10.1088/1475-7516/2015/03/010}{\emph{JCAP} {\bfseries
  1503} (2015) 010} [\href{https://arxiv.org/abs/1409.5135}{{\ttfamily
  1409.5135}}].

\bibitem{doi:10.1021/jp970984n}
D.~J. Wales and J.~P.~K. Doye, \emph{Global optimization by basin-hopping and
  the lowest energy structures of lennard-jones clusters containing up to 110
  atoms}, \href{https://doi.org/10.1021/jp970984n}{\emph{The Journal of
  Physical Chemistry A} {\bfseries 101} (1997) 5111}
  [\href{https://arxiv.org/abs/https://doi.org/10.1021/jp970984n}{{\ttfamily
  https://doi.org/10.1021/jp970984n}}].

\bibitem{Obied:2018sgi}
G.~Obied, H.~Ooguri, L.~Spodyneiko and C.~Vafa, \emph{{De Sitter Space and the
  Swampland}},  \href{https://arxiv.org/abs/1806.08362}{{\ttfamily
  1806.08362}}.

\bibitem{Agrawal:2018own}
P.~Agrawal, G.~Obied, P.~J. Steinhardt and C.~Vafa, \emph{{On the Cosmological
  Implications of the String Swampland}},
  \href{https://arxiv.org/abs/1806.09718}{{\ttfamily 1806.09718}}.

\bibitem{Baumann:2014nda}
D.~Baumann and L.~McAllister, \emph{{Inflation and String Theory}}, Cambridge
  Monographs on Mathematical Physics. Cambridge University Press, 2015,
  \href{https://doi.org/10.1017/CBO9781316105733}{10.1017/CBO9781316105733},
  [\href{https://arxiv.org/abs/1404.2601}{{\ttfamily 1404.2601}}].

\end{thebibliography}\endgroup
\bibliographystyle{JHEP}

\appendix
\section{Potential consistency}
\label{sec:potconst}
The potential construction presented here, similar to other DBM constructions \cite{Marsh:2013qca,Battefeld:2014qoa,Dias:2016slx,Dias:2017gva}, is not single-valued.
When the inflationary trajectory includes a loop in field space, DBM potentials are very unlikely to come back to the same value around the loop.
DBM potentials carry no global information and have no mechanism to enforce that the generated trajectory avoids this behavior.
We consider such trajectories unphysical and discard them.
In our simulations, we detect looping trajectories by constructing a hypercylinder\footnote{\textit{hypercylinder} meaning a hypersphere in directions orthogonal to the trajectory, and a line segment along the trajectory.} around each consecutive pair of points of the trajectory.
If another point of the trajectory is within any cylinder, we discard that realization. This check is $ \mathcal{O}(N^2) $, and trajectories of $N\approx 10^5$ points are not uncommon. In order to expedite the check, we only check the first and last tenths of the trajectory, roughly when $\eta_\perp$ is large and the radius of curvature is small.

The cylinders' radius is chosen to be nonzero to ensure there are no intersections within numerical accuracy, and that perturbations in the local neighborhood of the trajectory are in a smooth region. In all realizations presented here, we chose the cylinder radius to be $10^{-7}\Lambda_h$.

The chance for a self-intersection does not always decrease with the number of fields.
For example, here are the $\Lambda_h=0.4\Mpl$ realizations from dataset 6:
\begin{table}[H]
    \centering
    \begin{tabular}{|c|c|}
    \hline
    $N_f$ & \% self-intersect \\ \hline
    10 & 0.0 \\ \hline 
    40 & 10.1 \\ \hline
    80 & 13.5 \\ \hline
    \end{tabular}
\end{table}
Though the dimension of field space increases with higher $N_f$, there is on average only one approximately flat direction.
We chose a random direction for our initial velocity, so in higher-$N_f$ potentials, the chance the initial velocity points uphill increases.
Uphill velocities will slow and turn back, possibly entering the $10^{-7}\Lambda_h$ intersection radius.
\FloatBarrier
\section{Convergence}
In random potentials, observables do not have concrete predicted values, but rather distributions of values given by the model.
It is far from obvious that taking $\mathcal{O}(10^3)$ realizations per data point was a sufficient sampling of the underlying distribution in our potentials.
Below we show a dataset with $5000$ realizations, of which $2104$ successfully computed perturbations.

\begin{figure}[H]
    \centering
    \includegraphics[width=0.5\textwidth]{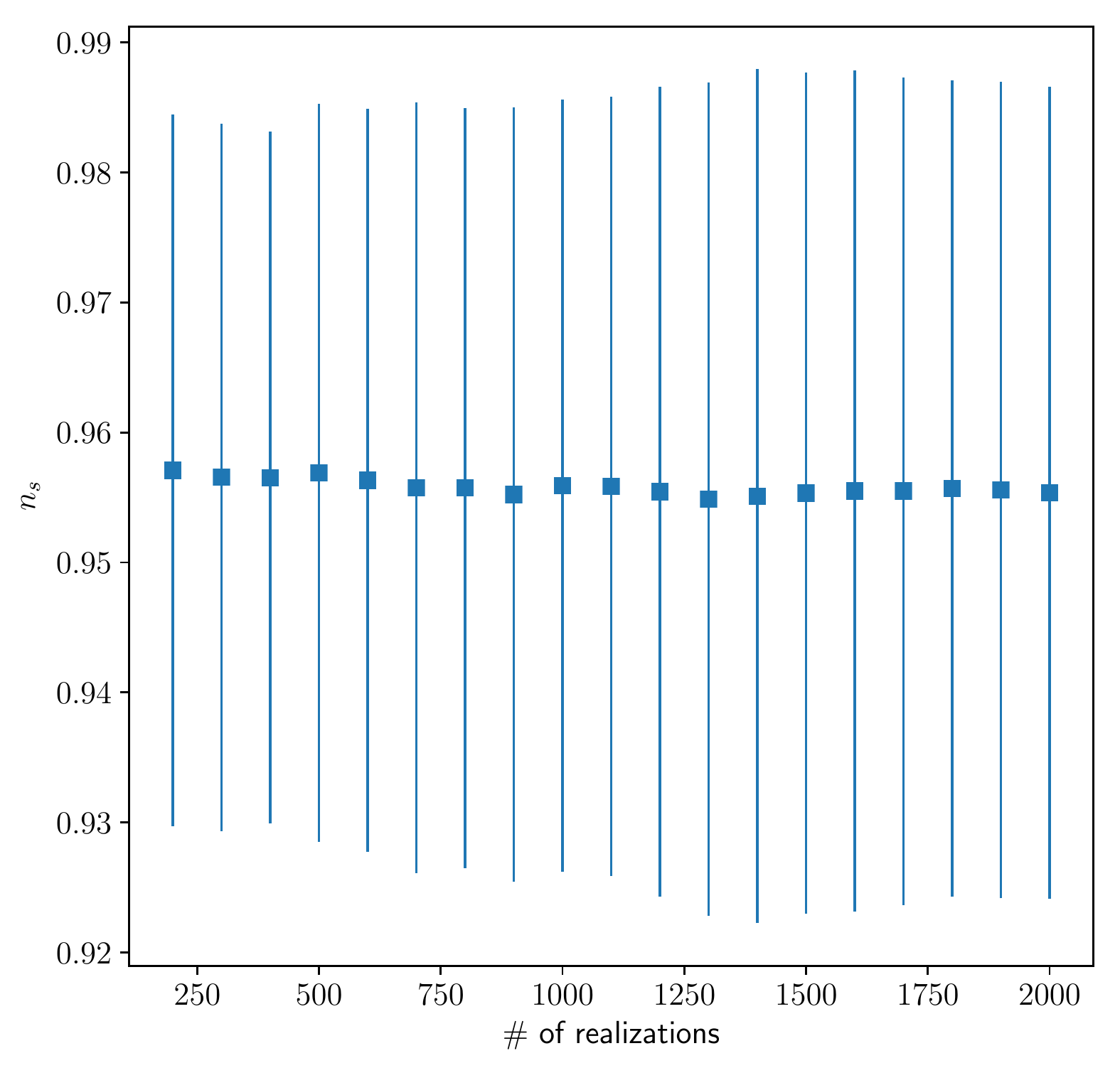}
    \caption{The predictions for $n_s$ from a dataset with $\sim 2000$ realizations, as we vary the number of included realizations in the average. The mean is well stabilized by $1000$ realizations, but the standard deviation may vary from the underlying standard deviation by several percent.}
\end{figure}

\section{Positive $\beta$}
\begin{figure}[H]
    \centering
    \includegraphics[width=0.9\textwidth]{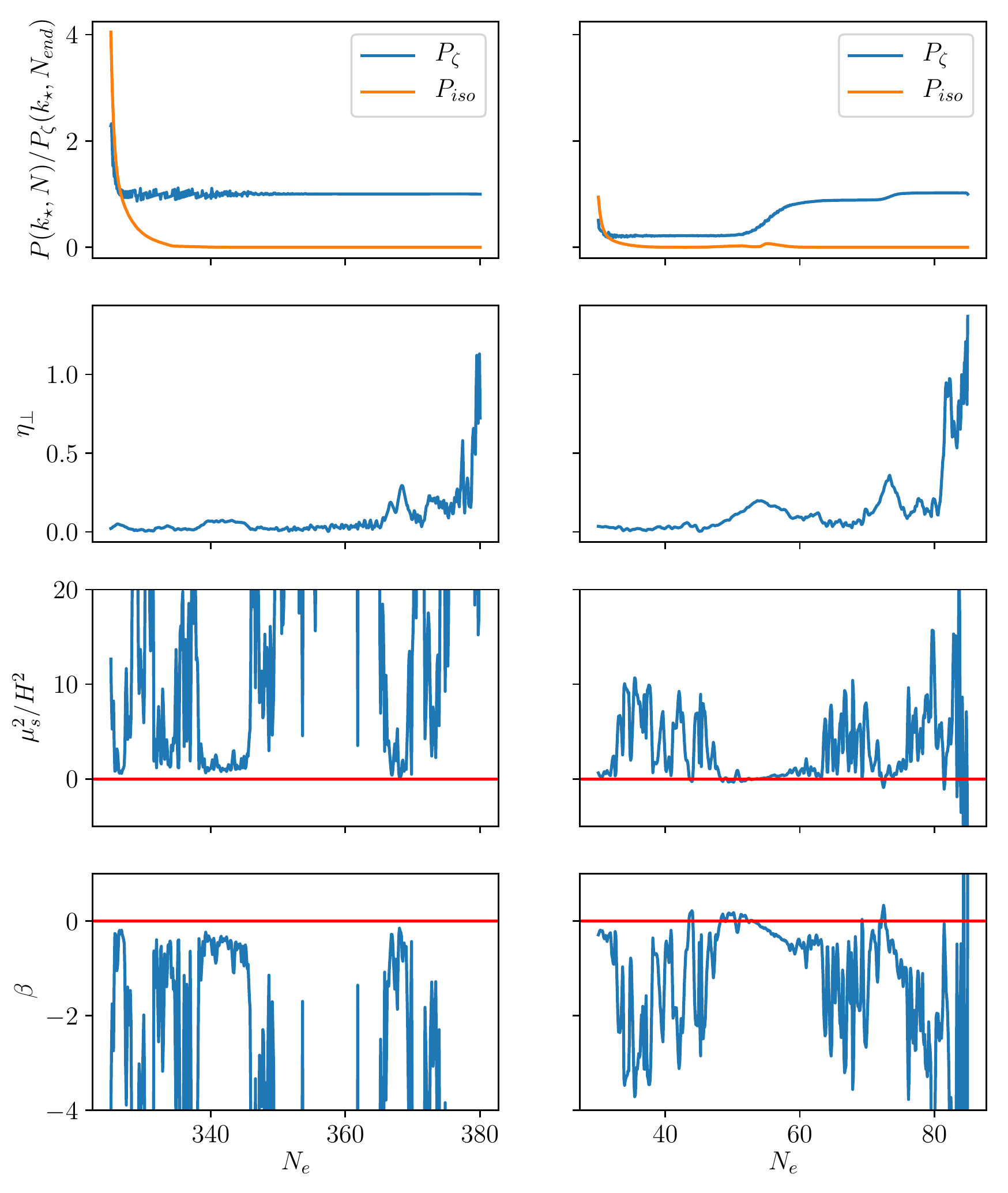}
    \caption{A typical 5-field realization (left) and an atypical one with large isocurvature (right). The mechanism described in section \ref{sec:perturbations} where the first isocurvature mode sources the adiabatic mode occurs here.}
    \label{fig:pos_beta}
\end{figure}

\end{document}